\documentclass[12pt]{article}
\usepackage{latexsym}
\usepackage{amsmath}
\usepackage{amssymb}
\usepackage{relsize}
\usepackage{geometry}
\usepackage{tensor}
\usepackage{physics} 
\usepackage{csquotes} 

\def\beqn{\begin{eqnarray}}
\def\eeqn{\end{eqnarray}}

\def\beq{\begin{equation}}
\def\eeq{\end{equation}}
\def\ba{\beq\new\begin{array}{c}}
\def\ea{\end{array}\eeq}

\def\Tr{{\rm Tr}}
\newcommand{\gsim}{\lower.7ex\hbox{$
\;\stackrel{\textstyle>}{\sim}\;$}}
\newcommand{\lsim}{\lower.7ex\hbox{$
\;\stackrel{\textstyle<}{\sim}\;$}}

\newcommand{\ntwo}{${\mathcal N}=2$ }

\newcommand{\ntwot}{${\mathcal N}= \left(2,2\right) $ }
\newcommand{\ntwoo}{${\mathcal N}= \left(0,2\right) $ }
\newcommand{\none}{${\mathcal N}=1$ }

\newcommand{\vp}{\varphi}
\newcommand{\pt}{\partial}

\newcommand{\qt}{\tilde q}
\renewcommand{\theequation}{\thesection.\arabic{equation}} 

\newcommand{\p}{\partial}
\newcommand{\wt}{\widetilde}
\newcommand{\ov}{\overline}
\newcommand{\mc}[1]{\mathcal{#1}}
\newcommand{\md}{\mathcal{D}}

\newcommand{\lgr}{\left\lgroup}
\newcommand{\rgr}{\right\rgroup}

\def\slashed#1{\setbox0=\hbox{$#1$}             
   \dimen0=\wd0                                 
   \setbox1=\hbox{/} \dimen1=\wd1               
   \ifdim\dimen0>\dimen1                        
      \rlap{\hbox to \dimen0{\hfil/\hfil}}      
      #1                                        
   \else                                        
      \rlap{\hbox to \dimen1{\hfil$#1$\hfil}}   
      /                                         
   \fi}                                        %

\newcommand{\AU}{\mc{A}^{\rm U(1)}}
\newcommand{\AN}{\mc{A}^\text{SU($N$)}}
\newcommand{\aU}{a^{\rm U(1)}}
\newcommand{\aN}{a^\text{SU($N$)}}
\newcommand{\baU}{\ov{a}{}^{\rm U(1)}}
\newcommand{\baN}{\ov{a}{}^\text{SU($N$)}}
\newcommand{\lU}{\lambda^{\rm U(1)}}
\newcommand{\lN}{\lambda^\text{SU($N$)}}

\newcommand{\nbar}{\ov{n}}

\usepackage{mathtools}
\usepackage{hyperref}
\usepackage{xcolor}

\begin{document}

\hypersetup{%
	colorlinks=false,
	linkbordercolor=blue,
	pdfborderstyle={0 0 0.1}%
}

\begin{titlepage}

\begin{flushright}
FTPI-MINN-17/07, UMN-TH-3623/17\\
\end{flushright}

\begin{center}
	\Large{{\bf Non-Abelian strings in \none supersymmetric QCD}}

\vspace{5mm}

{\large \bf E.~Ievlev$^{\,a,b}$ and  A.~Yung$^{\,a,b,c}$}
\end {center}

\begin{center}

$^{a}${\it National Research Center ``Kurchatov Institute'',
Petersburg Nuclear Physics Institute, Gatchina, St. Petersburg
188300, Russia}\\
$^{b}${\it  St. Petersburg State University,
 Universitetskaya nab., St. Petersburg 199034, Russia}\\
 $^c${\it  William I. Fine Theoretical Physics Institute,
University of Minnesota,
Minneapolis, MN 55455}\\
\end{center}

\vspace{1cm}

\begin{abstract}
Non-Abelian flux tubes (strings) are well studied in \ntwo supersymmetric QCD in (3+1) dimensions. In addition to
translational zero modes  they have also orientational moduli associated with rotations of their fluxes inside a non-Abelian group. The dynamics of the orientational moduli is described by the two dimensional CP$(N-1)$ model living on the world sheet of the non-Abelian string. In this paper we consider a deformation of \ntwo supersymmetric QCD with the U$(N)$ gauge group and $N_f=N$ quark flavors with a mass term $\mu$ of the adjoint matter. In the limit of large $\mu$ the theory flows to an \none supersymmetric QCD. We study the solution for the  non-Abelian string in this
limit and derive an effective theory on the string world sheet. The bosonic sector of this theory is still given by the CP$(N-1)$ model but its scale is exponentially small as compared to the scale of the four dimensional bulk theory in contrast to the \ntwo case where these scales are equal. We study also the fermionic sector of the world sheet theory. Upon the deformation the
non-Abelian string is no longer BPS and we show that the fermionic superorientational zero modes are all lifted. This leaves us with the pure bosonic CP$(N-1)$ model on the string world sheet in the limit of \none QCD.
We also discuss what happens to confined monopoles at large $\mu$.

\end{abstract}

\end{titlepage}

\newpage

\tableofcontents
\clearpage

%
%

\newpage

\section*{Introduction}  \label{introduction}
\addcontentsline{toc}{section}{Introduction}

The mechanism of confinement based on a  monopole condensation  \cite{mandelstam} was shown
to work \cite{SW1,SW2} in the
monopole vacua of \ntwo supersymmetric QCD. This confinement  is essentially
Abelian \cite{DS,HSZ,Strassler,VY}. Non-Abelian gauge group is broken down to an Abelian subgroup
by condensation of the adjoint scalars at a high scale, while the condensation of  monopoles occurs  at a much lower scale, in a low-energy Abelian theory.
Simultaneously, there occurs the
formation of confining flux tubes (strings) of the Abelian Abrikosov-Nielsen-Olesen (ANO) type \cite{ANO}.

On the other hand it is believed that confinement in the real world QCD is essentially non-Abelian. This motivates studies of possible non-Abelian generalizations of the confinement mechanism.
One important ingredient of such mechanism, namely the non-Abelian confining strings, was found in
 \ntwo supersymmetric QCD \cite{HT1,ABEKY,SYmon,HT2}, see also \cite{Trev,Jrev,SYrev,Trev2} for a review. These strings are formed in the quark vacua of \ntwo QCD with the U$(N)$ gauge group, and they
 give rise to the confinement of monopoles and the so called ''instead-of-confinement'' phase for quarks, see \cite{SYdualrev} for a review.

In much the same way as the real world QCD,
 \none supersymmetric QCD does not have  adjoint scalars.  Therefore it is believed to have an essentially non-Abelian dynamics. On the other hand, due to supersymmetry it is more tractable then non-supersymmetric QCD. One may hope that, starting
from \ntwo QCD and decoupling the adjoint scalars, one can arrive at a non-Abelian
regime. In particular, it was shown that the non-Abelian ''instead-of-confinement'' phase
survives in the limit where the adjoint matter (present in  \ntwo QCD)  decouples, see review \cite{SYdualrev} and references therein.

In this paper we make the next step  and study what happens to the non-Abelian confining strings upon decoupling of the adjoint matter. Namely, we consider a deformation of  \ntwo supersymmetric QCD with the U$(N)$ gauge group and $N_f=N$ quark flavors by a mass term $\mu$ of the adjoint matter. The $\mu$-deformation breaks the \ntwo supersymmetry and in the limit of large $\mu$ the theory flows to \none supersymmetric QCD.

In addition to the
translational zero modes typical for ANO strings, non-Abelian strings have  orientational moduli associated with rotations of their fluxes inside the non-Abelian SU$(N)$ group. The dynamics of the orientational moduli in \ntwo QCD is described by the two dimensional CP$(N-1)$ model living on the world sheet of the non-Abelian string. In this paper we study the solution for the  non-Abelian string  and derive an effective theory on the string world sheet in the
limit of large $\mu$.

Similar problem was  addressed in \cite{SYnone,Edalati,Tongd,SY02,SYhetN,BSYhet} where the $\mu$-deformation was considered in \ntwo supersymmetric QCD with the U$(N)$ gauge group and $N_f=N$ flavors of massless  quarks supplemented by the Fayet-Iliopoulos (FI) $D$-term.  In the limit of large $\mu$ this theory flows to
a theory which differs from \none QCD by the presence of the  FI term. In particular,
in this theory a scalar quark (squark) condensation is triggered by the FI $D$-term.

It was shown in the aforementioned papers that bosonic profile functions of the non-Abelian string stay intact
upon the $\mu$-deformation, while the fermionic zero modes are changed as compared to the ones in the \ntwo limit. The string remains BPS saturated and the world sheet theory becomes the heterotic CP$(N-1)$ model with \ntwoo supersymmetry \cite{Edalati,Tongd,SY02,BSYhet}. In this model, the supertranslational fermionic moduli interact with the superorientational ones. Large $N$ solution of the world sheet
model shows that \ntwoo supersymmetry  is spontaneously broken \cite{SYhetN}. The model has $N$ vacua corresponding to $N$ different non-Abelian strings and the discrete  $Z_{2N}$ symmetry is
spontaneously broken.

In this paper we consider the $\mu$-deformation of \ntwo QCD without a FI term in a quark vacuum. Squark condensate is determined by $\sqrt{\mu m}$, were $m$ is a quark mass. In the large $\mu$ limit the theory flows to  \none QCD in the quark vacuum. Non-Abelian strings cease to be BPS
saturated and both bosonic and fermionic profile of the string are modified.

We study solutions for the  non-Abelian string profile functions in the large $\mu$
limit and derive the effective theory on the string world sheet. The bosonic sector of this theory is still given by the CP$(N-1)$ model. The CP$(N-1)$ model is asymptotically free, and it
is determined by its scale $\Lambda_{CP}$ (position of the infra-red pole of the coupling constant). 
At small $\mu$ $\Lambda_{CP} = \Lambda_{{\cal N}=2}$, where $\Lambda_{{\cal N}=2}$ is the scale of
four dimensional \ntwo QCD, see, for example, review \cite{SYrev}.
We show that in the in the large $\mu$
limit $\Lambda_{CP}$ is exponentially small.
We also derive a potential in  two dimensional world sheet theory induced by quark mass differences.

 Next we study  the fermionic sector of the world sheet theory. Upon the $\mu$-deformation the
 fermionic superorientational zero modes are all lifted. This leaves us with the pure bosonic CP$(N-1)$ model on the string world sheet in the limit when the bulk theory becomes \none QCD. This ensures that the
 world sheet theory is in the Coulomb/confinement phase, at least at large $N$, see \cite{W79}.

We also address a question of what happen to the confined 't Hooft-Polyakov monopoles 
present in the \ntwo limit,
when we go to the large $\mu$. Studying the world sheet potential we show that 
confined monopoles seen in the world sheet theory
as kinks \cite{SYmon,HT2} become unstable at large $\mu$ if quark masses are not equal.  However, if 
 quarks have equal masses the confined monopoles survive in the limit of \none QCD.

The  paper is organized as follows.
In Sec.~\ref{sec:basics} we review our bulk theory, namely  $\mu$-deformed \ntwo supersymmetric QCD 
in the quark vacuum and calculate its perturbative  mass spectrum in the large $\mu$ limit. 
Sec.~\ref{sec:bosonic_solution} presents the non-Abelian bosonic string solution at
large $\mu$ for equal as well as unequal quark masses. The bosonic part
of the  effective world sheet theory is derived in Sec.~\ref{sec:effective_bosonic_action}.
First we consider the kinetic term and then discuss the world sheet potential which arises
when quark masses are not equal.
Moving on, in Sec.~\ref{sec:superorient-zero} we study the fermionic sector and derive 
superorientational zero modes, starting from the \ntwo limit and then moving to the $\mu$-deformed  case.
We show that all the superorientational zero modes are lifted upon the $\mu$-deformation.
In Sec~\ref{physics} we review the physics of the world sheet theory on the non-Abelian  string 
and discuss what happen to confined monopoles at large $\mu$.
Sec.~\ref{Conclusions} is devoted to our
 Conclusions. Details of the derivation of the fermion zero 
modes are presented in Appendices.

%
%

\section{Bulk theory \label{sec:basics}}
\setcounter{equation}{0}


In this section we review our four dimensional bulk theory, see review \cite{SYrev} for more details. The bulk theory  is $\mu$-deformed \ntwo supersymmetric QCD (SQCD) with the gauge group U$(N)=$SU$(N)\times$U(1).
The field content of the theory is as follows.
The \ntwo vector multiplet consists of the U(1) gauge field $A_\mu$ and SU($N$) gauge field $A^a_\mu$, complex scalar fields $a^{U(1)}$ and $a^a$ in the adjoint representation, and their  fermion superpartners ($\lambda^{1}_{\alpha}$, $\lambda^{2}_{\alpha}$) and ($\lambda^{1a}_{\alpha}$, $\lambda^{2a}_{\alpha}$). The adjoint index $a$ runs from 1 to $N^2 - 1$, while the spinorial index $\alpha = 1,2$.
The adjoint scalars and fermions $\lambda^{2}$ can be combined into the \none adjoint matter chiral multiplets $ \AU $ and $ \AN = \mc{A}^a T^a $, where $T^a$ are generators of the SU($N$) gauge group normalized as
$\Tr \left( T^a T^b \right) ~=~ (1/2) \, \delta^{ab}~$.

The matter sector consists of $N_f = N$ flavors of quark hypermultiplets in the fundamental representation and scalar components (squarks) $q^{kA}$ and $\wt{q}_{Ak}$, while the fermions are represented by $\psi^{kA}$ and $\wt{\psi}_{Ak}$. Here $A = 1,..,N$ is a flavor index and $k=1,..,N$ is a color index.

The  superpotential of \ntwo supersymmetric QCD reads
\begin{equation}
 \mc{W}_{\mc{N}=2} ~~=~~  \sqrt{2}\, \Bigl\{
	 \frac12 \wt{q}{}_A \AU q^A ~+~
	 \wt{q}{}_A \mc{A}^a T^a q^A \Bigr\}  ~+~
	 m_A\, \wt{q}{}_A q^A \ ,
\label{supN2}
\end{equation}
where we use the same notations for quark multiplets $ q^A $ and $ \wt{q}{}_A $ and their scalar components, while $m_A$ are quark masses.

The $\mu$-deformation is the mass term for the adjoint matter
\begin{equation}
 \mc{W}_{\mc{N}=1} ~~=~~ \sqrt{\frac{N}{2}}\,\frac{\mu_1}{2} \left(\mc{A}^{\rm U(1)}\right)^2  ~~+~~
	 \frac{\mu_2}{2} \left( \mc{A}^a \right)^2 ~,
\label{none_superpotential_general}
\end{equation}
which breaks \ntwo supersymmetry down to \none.

In the special case when
\begin{equation}
\mu ~~\equiv~~\mu_2 ~~=~~ \mu_1 \sqrt{\frac{2}{N}}  ,
\label{mu_condition}
\end{equation}
superpotential  \eqref{none_superpotential_general} becomes a single trace operator
\begin{equation}
\mc{W}_{\mc{N}=1} = \mu \Tr (\Phi^2) \,
\end{equation}
where we defined a scalar adjoint matrix as
\beq
\Phi = \frac12 a^{U(1)} + T^a\, a^a.
\label{Phidef}
\eeq

We will consider bulk QCD in the limit of large $\mu_1$ and $\mu_2$, when the
adjoint matter decouples and the theory becomes \none QCD.
 Integrating  out the adjoint matter in a sum of superpotentials (\ref{supN2}) and (\ref{none_superpotential_general}) we get a quark superpotential of our $\mu$-deformed
 bulk theory
\begin{equation}
 \mc{W} (q, \tilde{q}) = - \frac{1}{2\mu_2} \left[ (\tilde{q}_A q^B)(\tilde{q}_B q^A) - \frac{\alpha}{N} (\tilde{q}_A q^A)^2 \right] + m_A (\tilde{q}_A q^A) \,,
\label{superpotential-adjoints_integrated}
\end{equation}
where
\begin{equation}
\alpha= 1- \sqrt{\frac{N}{2}} \frac{\mu_2}{\mu_1} \,.
\label{alpha}
\end{equation}
In the case of single trace deformation (\ref{mu_condition}) $\alpha =0$.

The bosonic action of the theory is %
 \footnote{From here further on we use a  Euclidean notation, that is $F_{\mu\nu}^2 = 2F_{0i}^2 + F_{ij}^2$, $\, (\partial_\mu a)^2 = (\partial_0 a)^2 +(\partial_i a)^2$, etc.  Furthermore, the sigma-matrices are defined as $\sigma^{\alpha\dot{\alpha}}=(1,-i\vec{\tau})$,
	$\bar{\sigma}_{\dot{\alpha}\alpha}=(1,i\vec{\tau})$. Lowering and raising
	of spinor indices are performed by  virtue of an anti-symmetric tensor
	defined as $\varepsilon_{12}=\varepsilon_{\dot{1}\dot{2}}=1$,
	$\varepsilon^{12}=\varepsilon^{\dot{1}\dot{2}}=-1$.
	The same raising and lowering convention applies to the flavor SU($N$)
	indices $f$, $g$, etc. }
\begin{align}
\label{theory}
S_{\rm bos} ~~=~~ & \int d^4 x
\lgr
\frac{1}{2g_2^2}\Tr \left(F_{\mu\nu}^\text{SU($N$)}\right)^2  ~+~
\frac{1}{4g_1^2} \left(F_{\mu\nu}^{\rm U(1)}\right)^2 ~+~
\right.
\\[2mm]
\notag
&
\phantom{int d^4 x \lgr\right.}
\left.
\left| \nabla_\mu q^A \right|^2 ~+~ \left|\nabla_\mu \ov{\wt{q}}{}^A \right|^2
~+~ V(q^A, \wt{q}_A)
\rgr .
\end{align}
Here $ \nabla_\mu $ is a covariant derivative
\beq
\nabla_\mu  ~~=~~ \p_\mu ~-~ \frac{i}{2}\,A^{\rm U(1)}_\mu ~-~ i\, A_\mu^a T^a~,
\label{covder}
\eeq
while the scalar potential $V(q^A, \wt{q}_A)$ is a sum of the $D$-term and $F$-term potentials,
\beq
V(q^A, \wt{q}_A)=V_D (q^A, \wt{q}_A) + V_F (q^A, \wt{q}_A).
\label{scpot}
\eeq
The $D$-term potential reads
\begin{equation}
	V_D ~~=~~ \frac{g_2^2}{2}\left(\bar{q}_A T^a q^A - \qt_A T^a \bar{\qt}^A\right)^2 ~+~
				\frac{g_1^2}{8} \left(|q^A|^2 - |\qt^A|^2\right)^2,
\label{potential:3+1:D:original}
\end{equation}
while the $F$ term potential is determined by superpotential (\ref{superpotential-adjoints_integrated}). It has the form
\begin{equation}
	\begin{aligned}
		V_F ~~=~~ \frac{1}{|\mu_2|^2}\,
		&\Bigg\{ (\bar{q}_A q^B)\, \Big[ (\bar{q}_C \bar{\qt}^A) - \frac{\bar{\alpha}}{N} \delta^A_C (\bar{q}_F \bar{\qt}^F) - \bar{\mu_2}\bar{m}_A\delta_C^A \Big]	\\
		&\times \Big[ (\qt_B q^C) - \frac{\alpha}{N} \delta_B^C (\qt_F q^F) - \mu_2 m_B\delta_B^C \Big]\\
		&+ (\qt_A \bar{\qt}^B)\, \Big[ (\bar{q}_B \bar{\qt}^C) - \frac{\bar{\alpha}}{N} \delta_B^C (\bar{q}_F \bar{\qt}^F) - \bar{\mu_2}\bar{m}_B\delta_B^C \Big]	\\
		&\times \Big[ (\qt_C q^A) - \frac{\alpha}{N} \delta_C^A (\qt_F q^F) - \mu_2 m_A\delta_C^A \Big]
		\Bigg\} \,.
	\end{aligned}
\label{potential:3+1:F:original}
\end{equation}

In this paper we will consider the vacuum (zero of the potential (\ref{scpot})) where the maximal possible number of quark flavors equal to $N$ condense (the so called $r=N$ vacuum, where $r$ is the number of condensed squark flavors at weak coupling, see \cite{SYdualrev} for  a review). In this vacuum squark VEVs are given by
\beq
\langle q^{kA} \rangle ~~=~~ \langle \ov{\wt{q}}{}^{kA} \rangle = \frac1{\sqrt{2}}
\begin{pmatrix}
\sqrt{\xi_1}  &   0  &  ... \\
... &  ... &  ... \\
... &   0  &  \sqrt{\xi_N}
\end{pmatrix}  , \\
\label{qVEVdifferent}
\eeq
where we wrote down the squark field as an $N\times N$ matrix in color and flavor indices, and the parameters $\xi_A$ are defined as
\begin{equation}
	\xi_A ~~=~~ 2\left( \sqrt{\frac{2}{N}} \mu_1 \widehat{m} ~+~ \mu_2 (m_A - \widehat{m}) \right) ,
	\label{xi-general}
\end{equation}
while
\begin{equation}
	\widehat{m} ~~=~~ \frac{1}{N} \sum_{A=1}^{N} m_A .
\label{averagemass}
\end{equation}

For single trace deformation (\ref{mu_condition}) expressions for the parameters $\xi_A$ simplify:
\beq
\xi_A= 2 \,\mu_2 m_A
\label{xisingletrace}
\eeq

In this paper we will mostly consider the non-Abelian limit when all quark masses are equal,
\begin{equation}
m_1=m_2=...=m_N \equiv m ,
\label{equalmasses}
\end{equation}
so that the parameters $\xi_A$ degenerate, $\xi_A\equiv\xi$, and the squark VEVs become
\beq
\langle q^{kA} \rangle ~~=~~ \langle \ov{\wt{q}}{}^{kA} \rangle =
\sqrt{\frac{\xi}{2}}
\begin{pmatrix}
1  &   0  &  ... \\
... &  ... &  ... \\
... &   0  &  1
\end{pmatrix}
\label{qVEV}
\eeq

Note that if we take the limit $\mu\to\infty$ (keeping the quark masses fixed) the parameters
$\xi\sim \mu m$ also go to infinity, and our quark vacuum becomes a run-away vacuum (all the $r$ vacua
with the non-zero $r$ become run-away vacua). In this case \none QCD is a theory with only $N$ vacua
which originate from $N$ monopole vacua ($r=0$ vacua) of \ntwo QCD.

Here we define \none QCD in a different way. By taking the limit of large $\mu$ we make the quark masses
small so that the product $\mu m$ (and the quark VEVs) are fixed,
\beq
\mu \to \infty, \qquad m\to 0, \qquad \mu\,m = {\rm fixed}.
\label{mutoinfty}
\eeq
 This way we keep track of all the $r$ vacua present in \ntwo QCD. In this paper we will study non-Abelian strings particularly in the
$r=N$ quark vacuum (\ref{qVEV}) assuming the limit of large $\mu$ when the bulk theory flows to the
generalized \none QCD defined above.

In order to keep our bulk theory at weak coupling we assume that the squark VEVs are large as compared with the scale $\Lambda_{{\cal N}=1}$ of the SU$(N)$ sector of \none QCD. Namely, we assume that
\beq
\sqrt{\mu m} \gg \Lambda_{{\cal N}=1}.
\label{weakcoupl}
\eeq

Squark VEVs (\ref{qVEV}) result in  a spontaneous
breaking of both gauge and flavor SU($N$)'s.
The diagonal global SU($N$) survives, however,
\beq
{\rm U}(N)_{\rm gauge}\times {\rm SU}(N)_{\rm flavor}
\to {\rm SU}(N)_{C+F}\,.
\label{c+f}
\eeq
A color-flavor locking takes place in the vacuum. This fact  leads to an emergence of non-Abelian strings, see \cite{SYrev} for a review.

Let us briefly summarize a perturbative spectrum of our bulk theory in the large $\mu$ limit,
cf. \cite{SYrev}. Consider for simplicity the case of equal quark masses. The U$(N)$ gauge group is completely Higgsed and  the masses of the gauge bosons are
\begin{equation}
	m_G^{SU(N)} ~~=~~ g_2|\sqrt{\xi}| ~
\label{massGSUN}
\end{equation}
for the SU$(N)$ gauge bosons and
\begin{equation}
	m_G^{U(1)} ~~=~~ g_1 \sqrt{\frac{N}{2}} |\sqrt{\xi}|~
\label{massGU1}
\end{equation}
for the U(1) one. Below we also assume that the gauge boson masses are of the same order,
\beq
m_G^{U(1)} \sim m_G^{SU(N)} \equiv m_G
\label{massG}
\eeq

Extracting a quark mass matrix from potentials (\ref{potential:3+1:D:original}), (\ref{potential:3+1:F:original}) we find that out of $4N^2$ real degrees of freedom of the $q^{kA}$ and $\ov{\wt{q}}{}^{kA}$ squarks $N^2$ phases are eaten by the Higgs mechanism, $(N^2-1)$ real squarks have mass (\ref{massGSUN}), while one real squark has mass (\ref{massGU1}). These squarks
are scalar superpartners of the SU$(N)$ and U$(1)$ gauge bosons in massive vector \none supermultiplets, respectively.

Other  $2N^2$ squarks become much lighter in the large $\mu$ limit. The masses of $2(N^2-1)$ of
them forming the adjoint representation of the global color-flavor SU$_{C+F}(N)$ (\ref{c+f}) are given by
\beq
m_L^{SU(N)} =\left|\frac{\xi}{\mu_2}\right|,
\label{massLSUN}
\eeq
while  two real SU$_{C+F}(N)$ color-flavor singlets have mass
\beq
m_L^{U(1)} =\sqrt{\frac{N}{2}}\,\left|\frac{\xi}{\mu_1}\right|,
\label{massLU1}
\eeq

If $\mu_2$ and $\mu_1$ are of the same order (more exactly, we assume  below that  $\alpha = {\rm const}$, see \eqref{alpha}), then
\beq
 m_L^{U(1)} \sim m_L^{SU(N)} \equiv m_L   \sim m \ll m_G.
\label{masshierarchy}
\eeq
Below we will heavily use this mass hierarchy of the perturbative spectrum.

In particular, in the limit (\ref{mutoinfty}) $m_L\to 0$, and $2N^2$ squarks become massless.
This reflects the presence of the Higgs branch which develops  in this limit. The presence of
massless scalars developing VEVs makes the string solution ill-defined \cite{PennRubak,Y99}, see also next section.
Below we use the $\mu$-deformed \ntwo QCD at large $\mu$ as an infra-red (IR) regularization of \none
QCD. At large but finite $\mu$ the Higgs branch present in \none QCD is lifted and the IR
divergences are regularized, cf. \cite{EvlY}.

%
%

\section{Non-Abelian strings}
\label{sec:bosonic_solution}
\setcounter{equation}{0}

In this section we derive a vortex solution, assuming the equal quark mass limit (\ref{equalmasses}). First we review a general {\it ansatz} for the non-Abelian string and
present equations for string profile functions. Then we  solve these equations
assuming the mass hierarchy \eqref{masshierarchy} in the large $\mu$ limit.

\subsection{Equations of motion}

 We consider a static string stretched along the $x_3$ axis so that the corresponding profile functions depend only on coordinates in the $(x_1,x_2)$ plane. Closely following the strategy developed for \ntwo supersymmetic QCD (see review \cite{SYrev}) we first assume that
only those squark fields which develop VEVs have non-trivial profile functions in a string solution. Therefore we set
\begin{equation}
q^{kA}=\bar{\tilde{q}}^{kA}=\frac1{\sqrt{2}}\vp^{kA}  .
\label{q-ansatz}
\end{equation}
and look for the string solutions using the following {\em ansatz} \cite{ABEKY,SYmon,SYrev}:
\begin{equation}
	\begin{aligned}
		\varphi & ~~=~~
		\phi_2 ~+~ n\nbar\, \bigl( \phi_1 ~-~ \phi_2 \bigr)
		\\[2mm]
		&
		~~=~~
		\frac{1}{N}\bigl( \phi_1 ~+~ (N-1)\phi_2 \bigr)
		~+~ \bigl( \phi_1 ~-~ \phi_2 \bigr)
		\lgr n\nbar ~-~ 1/N \rgr \,,
		\\[2mm]
		A_i^\text{SU($N$)} & ~~=~~ \varepsilon_{ij}\, \frac{x^j}{r^2}\, f_{W}(r)
		\lgr n\nbar ~-~ 1/N \rgr\,,
		\\[2mm]
		A_i^{\rm U(1)} & ~~=~~ \frac{2}{N}\varepsilon_{ij}\, \frac{x^j}{r^2}\, f(r)~,
	\end{aligned}
\label{string-solution}
\end{equation}
where the index $i$ runs over 1, 2. The profile functions $ \phi_1(r) $ and $ \phi_2(r) $ determine the profiles of the squarks
in the plane orthogonal to the string at rest, while $ f(r) $ and $ f_{W}(r) $ are the profiles of the gauge
fields. The profile functions  depend on the distance $r$ from a given point $x^i$
to the center of the string $x^i_0$ in the $(x_1,x_2)$ plane.

Here we have also introduced the orientational complex  vector $n^l$, $l=1,...,N$, subject to the condition
\begin{equation}
\ov{n}{}_l \cdot n^l ~~=~~ 1 \, .
\end{equation}
 Vector $n^l$ parametrizes the orientational modes of the non-Abelian vortex string. It arises due to a possibility to rotate a given particular string solution with respect to the unbroken color-flavor global group SU$(N)_{C+F}$, see (\ref{c+f}).

Boundary conditions for the gauge and scalar profile functions are		
\begin{align}
\label{boundary}
\phi_1(0) & ~~=~~  0\text,                   & \phi_2(0) & ~~\neq~~ 0\text,  &
\phi_1(\infty) & ~~=~~ \sqrt{\xi} \text,     & \phi_2(\infty) & ~~=~~ \sqrt{\xi}\text, \\
\notag
f_{W}(0) & ~~=~~ 1\text,                   & f(0) & ~~=~~ 1\text,   &
f_{W}(\infty) & ~~=~~ 0 \text,            &  f(\infty) & ~~=~~ 0\text.
\end{align}

  Substituting {\it ansatz} (\ref{string-solution}) into action \eqref{theory} we get an energy functional (tension of the string):
\begin{multline}
	T ~~=~~ 2\pi \int rdr \Bigg( \frac{2}{g_1^2 N^2} \frac{f'^2}{r^2} ~+~ \frac{N-1}{N}\frac{1}{g_2^2} \frac{f_W'^2}{r^2} ~+~ \phi_1'^2 ~+~ (N-1)\phi_2'^2 \\
	~+~ \frac{1}{N^2} \frac{\left[f + (N-1)f_W\right]^2}{r^2}\phi_1^2 ~+~ \frac{N-1}{N^2} \frac{\left[f - f_W\right]^2}{r^2}\phi_2^2
	~+~ V(\phi_1, \phi_2) ,
	\Bigg)
\label{tenfunct}
\end{multline}
where the potential $V(\phi_1, \phi_2)$ is
\begin{multline}
	V(\phi_1, \phi_2) ~~=~~ \frac{1}{4|\mu_2|^2} \Bigg(
	\phi_1^2 \left[\phi_1^2 ~+~ \frac{\alpha}{N}(\phi_1^2 + (N-1)\phi_2^2) - 2\mu_2 m  \right]^2 \\
	~+~ (N-1)\phi_2^2 \left[\phi_2^2 ~+~ \frac{\alpha}{N}(\phi_1^2 + (N-1)\phi_2^2) - 2\mu_2 m  \right]^2
	\Bigg)
\label{potphi}
\end{multline}
and we assume that $\mu_2 m$ is real \footnote{If it is in fact a complex quantity, we should modify relation
\eqref{q-ansatz} inserting there the phase of $\mu_2 m$.}.

String tension functional \eqref{tenfunct}) gives  equations for the profile functions. We get
\begin{align}
\label{profileeqs}
	f'' ~-~ \frac{f'}{r} ~-~ \frac{g_1^2}{2}(f + (N-1)f_W)\phi_1^2 ~-~ (N-1) \frac{g_1^2}{2} (f - f_W) \phi_2^2 &~~=~~ 0\\
\notag
	f_W'' ~-~ \frac{f_W'}{r} ~-~ \frac{g_2^2}{N}(f + (N-1)f_W)\phi_1^2 ~+~ \frac{g_2^2}{N} (f - f_W) \phi_2^2 &~~=~~ 0 \\
\notag
	\phi_1'' ~+~ \frac{\phi_1'}{r} ~-~ \frac{1}{N^2}\frac{(f + (N-1)f_W)^2}{r^2}\phi_1 ~-~ \frac{1}{2} \frac{\p V}{\p \phi_1} &~~=~~ 0 \\
\notag
	\phi_2'' ~+~ \frac{\phi_2'}{r} ~-~ \frac{1}{N^2}\frac{(f - f_W)^2}{r^2}\phi_2 ~-~ \frac{1}{2(N-1)} \frac{\p V}{\p \phi_2} &~~=~~ 0
\end{align}
These equations are of the second order rather than the first order. This is because  our string is not  BPS saturated. Note, that for a BPS string the masses of the scalars forming the string are equal to masses of the gauge bosons \eqref{massGSUN} and \eqref{massGU1}, see \cite{SYrev}.
For our $\mu$-deformed theory this is not the case. Masses of singlet and adjoint scalars in the scalar matrix $\varphi^{kA}$ in \eqref{q-ansatz} are given by \eqref{massLSUN} and \eqref{massLU1}, and in the large $\mu$ limit they are much smaller than the masses of gauge bosons.
In particular, as we mentioned already, in the limit \eqref{mutoinfty} $m_L\to 0$, and our $\mu$-deformed theory develops a Higgs branch.

\subsection{String profile functions}
\label{sec:profile}

It is quite often that supersymmetric gauge theories have Higgs branches. These
are flat directions of the scalar potential on which charged scalar fields can
develop  VEVs breaking the gauge symmetry. In
many instances this breaking provides topological reasons behind formation
of  vortex strings. A dynamical side of the problem of the vortex string
formation in theories with Higgs branches was addressed in \cite{PennRubak,Y99,EvlY}. A
priori it is not clear at all whether or not stable string solutions exist in
this class of theories. The fact is that a theory with a Higgs branch
represents a limiting case of type I superconductor with vanishing Higgs
mass. In particular, it was shown in \cite{PennRubak} that infinitely long strings cannot be
formed in this case due to infrared divergences.

Later this problem was studied in \cite{Y99,EvlY}. It was shown that vortices on Higgs branches become logarithmically ''thick'' due to the presence of massless scalars in the bulk. Still,
they are well defined if IR divergences are regularized. One way of regularization is to
consider a vortex string of the finite length $L$ \cite{Y99}. This setup is typical for the confinement problem. It was shown in \cite{Y99} that confining potential between heavy trial charges becomes nonlinear,
\beq
V(L) \sim \frac{L}{\log{L}} ,
\eeq
in theories with Higgs branches.

Another way of IR regularization is to lift the Higgs branch so that scalar fields forming the string have small but non-zero masses $m_L$, cf. \cite{EvlY}. We use this approach here, see
Eqs. \eqref{massLSUN} and \eqref{massLU1}, assuming that $\mu$ is large but finite.

To the leading order in $\log{m_G/m_L}$ the vortex solution has the following structure in the
$(x_1,x_2)$ plane \cite{Y99}. The gauge fields are localized inside the core region of the radius
$R_g$ and almost zero outside this region\footnote{We will determine $R_g$ shortly.}. In contrast, scalar profiles are almost constant inside the core. In particular, the $\phi_1$ profile function associated with winding of the vortex is almost zero
inside the core (see \eqref{boundary}),
\begin{equation}
\begin{aligned}
\phi_1 &~~\approx~~ 0 \\
\phi_2 &~~\approx~~ (1-c)\sqrt{\xi},
\end{aligned}
\label{string:boson:large_mu:dm=0:small_r:phi}
\end{equation}
where $c$ is a constant to be determined.

Then the two first equations for gauge profile functions in \eqref{profileeqs} have solutions
\begin{equation}
	f ~~=~~ f_W ~~\approx~~ 1 ~-~ \frac{r^2}{R_g^2}
\label{string:boson:large_mu:dm=0:small_r:f}
\end{equation}
inside the core.

Outside the core in a logarithmically wide region
\beq
1/m_G \lesssim r \lesssim 1/m_L
\label{logregion}
\eeq
gauge fields are almost zero and two last equations in \eqref{profileeqs} reduce to the equations for free massless scalars. Their solutions have a logarithmic form
\begin{equation}
	\begin{aligned}
		\phi_1 &~~\approx~~ \sqrt{\xi} \left(1 ~-~ \frac{\ln \displaystyle\frac{1}{r m_L}}{\ln \displaystyle\frac{1}{m_L R_g}} \right) \,, \\
		\phi_2 &~~\approx~~ \sqrt{\xi} \left(1 ~-~ c\cdot\frac{\ln \displaystyle\frac{1}{r m_L}}{\ln \displaystyle\frac{1}{m_L R_g}} \right) \,,
	\end{aligned}
\end{equation}
where the normalization is fixed by matching with the behavior inside the core \eqref{string:boson:large_mu:dm=0:small_r:phi} and with the boundary conditions at infinity
\eqref{boundary}.

In the region of very large $r$, $r\gg 1/m_L$, the scalar fields exponentially approach  their
VEVs ($\sim \exp{-m_Lr}$), see \eqref{boundary}).
This region gives a negligible contribution to the string tension and a particular form of the scalar potential \eqref{potphi} is not important.

Upon a substitution of the above  solution into tension functional \eqref{tenfunct} one arrives at
\begin{equation}
	T ~~\approx~~ \frac{{\rm const}}{ R^2_g}\left(\frac2{g_1^2 N^2} + \frac{N-1}{g_2^2N}\right)+ \frac{2\pi|\xi|}{\ln \displaystyle\frac{1}{R_g m_L}} \left[ 1 ~+~ (N-1)\,c^2\right],
\label{T}
\end{equation}
where the first term comes from the gauge fields inside the core while the second term is produced by
the logarithmic integral over the region \eqref{logregion} coming from the kinetic terms of scalars.

Minimization of this expression with respect to the constant $c$ yields
\begin{equation}
	c ~=~ 0 ,
\end{equation}
so that the profile function $\phi_2$ does not depend on $r$ and is given by its VEV $\sqrt{\xi}$.

Minimizing \eqref{T} with respect to $R_g$ we find
\beq
R_g \sim \frac{\rm{const}}{m_G} \ln{\frac{m_G}{m_L}}.
\label{Rg}
\eeq

The solutions for string profile functions in the intermediate region \eqref{logregion} becomes
\begin{equation}
	\begin{aligned}
		\phi_1 &~~\approx~~ \sqrt{\xi} \left(1 ~-~ \frac{\ln \displaystyle\frac{1}{r m_L}}{\ln \displaystyle\frac{m_G}{m_L}} \right) \,, \\
		\phi_2 &~~\approx~~ \sqrt{\xi} \,, \\
		f &~~\approx~~ f_W ~~\approx~~ 0 \, ,
	\end{aligned}
\label{string:boson:large_mu:dm=0:intermediate}
\end{equation}

while  the final result for the tension of a non-Abelian string takes the form
\beq
T = \frac{2\pi|\xi|}{\ln \displaystyle\frac{m_G}{m_L}} +\cdots,
\label{ten}
\eeq
where corrections are suppressed by powers of large logarithm $\log{m_G/m_L}$.
The leading term here comes from quark kinetic energy ($(\phi_1')^2$) integrated over
intermediate region \eqref{logregion}, see \eqref{tenfunct}.
Note, that the logarithmic suppression of the string tension is not specific for non-Abelian strings. Similar expression was found for the ANO string on a Higgs branch \cite{Y99,EvlY}.

\subsection{Non-equal quark masses}

In this section we relax condition \eqref{equalmasses} and consider a string solution
assuming that quark mass differences are small,
\beq
\Delta m_{AB}=m_A-m_B \ll \widehat{m},
\label{smallmassdiff}
\eeq
where $\widehat{m}$ is the average quark mass, \eqref{averagemass}.

Non-equal quark masses break color-flavor symmetry \eqref{c+f} down to U$(1)^N$, so the orientational modes of the non-Abelian string are no longer zero modes. They become quasizero
modes in the approximation of small quark mass differences \eqref{smallmassdiff}, cf. \cite{SYrev}. In fact, in Sect. \ref{sec:worldsheetpot} we will derive a shallow world sheet potential
with $N$   extreme points associated with $Z_N$ strings.

Now we generalize the {\it ansatz} for the string solution \eqref{string-solution} as follows. First we
set an orientational vector
\beq
n^l = \delta^{lA_0}, \qquad A_0=1,...,N
\label{nA0}
\eeq
separating the $A_0$-th $Z_N$ string (the string associated with the winding of $A_0$ squark flavor, see \cite{SYrev}).

We expect that, much in the same way as for the equal quark masses case, the main contribution
to the string tension comes from logarithmically wide intermediate region \eqref{logregion},
while the string core does not contribute to the leading order. Then taking into account \eqref{smallmassdiff} we can neglect mass differences of
different gauge bosons setting
\beq
m_G \approx g_2\sqrt{|\widehat{\xi}|},
\label{averagemG}
\eeq
where
\begin{equation}
	\widehat{\xi} ~~=~~ \frac{1}{N} \sum_{A=1}^{N} \xi_A .
\label{averagexi}
\end{equation}

In this approximation we can use the same {\it ansatz} for the gauge fields as for the case of equal quark masses, see two last equations in \eqref{string-solution} with $n^l$ from \eqref{nA0}. Gauge fields are still parametrized by only two gauge profile functions $f(r)$ and $f_W$, which are non-zero inside the string core determined by $m_G$ \eqref{averagemG}.

The {\it ansatz} for the squark fields in \eqref{string-solution} is generalized as follows
\begin{equation}
	\begin{aligned}
		\vp &~~=~~
		\begin{pmatrix}
		\phi_1(r)	&	0		&	\ldots	&	0			&	\ldots	&	0	 \\
			0		&\phi_2(r)	&	\ldots	&	0			&	\ldots	&	0	 \\
		\ldots		&\ldots		&	\ldots	&	\ldots		&	\ldots	&\ldots	 \\
			0		&	0		&	\ldots	&	\phi_{A_0}(r)	&	\ldots	&	0	 \\
		\ldots		&\ldots		&	\ldots	&	\ldots		&	\ldots	&\ldots	 \\
			0		&	0		&	\ldots	&	0			&	\ldots	&\phi_N(r)	 \\
		\end{pmatrix}	
	\end{aligned},
\label{ansatz-dm_neq_0-n=1}
\end{equation}
where we introduce the profile functions $\phi_1,...,\phi_N$ for non-winding flavors $A\neq A_0$
while the profile function $\phi_{A_0}$ is associated with $A_0$-th winding flavor.

Boundary conditions for gauge profile functions are the same as in \eqref{boundary} while for
quarks we require
\begin{align}
	\phi_{A_0}(0) &~=~ 0\\
	\phi_A(\infty) &~=~ \sqrt{\xi_A}, \qquad A=1,...,N ,
\end{align}
where $\xi_A$ are given by \eqref{xi-general}.

Equations for the profile functions now read
\begin{align}
f'' ~-~ \frac{f'}{r} ~-~ \frac{g_1^2}{2}(f + (N-1)f_W)\phi_{A_0}^2 ~-~  \frac{g_1^2}{2} (f - f_W) \sum_{A \neq A_0}  \phi_A^2 &~~=~~ 0\\
\notag
f_W'' ~-~ \frac{f_W'}{r} ~-~ \frac{g_2^2}{N}(f + (N-1)f_W)\phi_{A_0}^2 ~+~ \frac{g_2^2}{N(N-1)}
(f - f_W)\sum_{A \neq A_0}  \phi_A^2 &~~=~~ 0 \\
\notag
\phi_{A_0}'' ~+~ \frac{\phi_{A_0}'}{r} ~-~ \frac{1}{N^2}\frac{(f + (N-1)f_W)^2}{r^2}\phi_{A_0} ~-~ \frac{1}{2} \frac{\p V}{\p \phi_{A_0}} &~~=~~ 0 \\
\notag
\phi_A'' ~+~ \frac{\phi_A'}{r} ~-~ \frac{1}{N^2}\frac{(f - f_W)^2}{r^2}\phi_A ~-~ \frac{1}{2} \frac{\p V}{\p \phi_A} &~~=~~ 0 ,\ A \neq A_0.
\end{align}

Solving these equations in much the same way as we did in the previous subsection we get
\beq
\phi_{A} ~~\approx~~ \sqrt{\xi_A}, \qquad A\neq A_0.
\label{phiA}
\eeq
Moreover,  gauge profile functions are determined by \eqref{string:boson:large_mu:dm=0:small_r:f}
inside the core, while they are zero outside it. Here the size of the core is still given by \eqref{Rg}.

In much the same way as for the case of equal quark masses the profile function $\phi_1$ is
almost zero inside the core and is given by
\beq
\phi_{A_0} ~~\approx~~ \sqrt{\xi_{A_0}} \left(1 ~-~ \frac{\ln \displaystyle\frac{1}{r m_L}}{\ln \displaystyle\frac{m_G}{ m_L}} \right) \,
\label{phi1massdiff}
\eeq
in the region \eqref{logregion} of intermediate $r$.

The results for the string tensions of  $Z_N$ strings have the form
\beq
T_{A_0} ~~=~~ \frac{2\pi |\xi_{A_0}|}{\ln \displaystyle\frac{m_G}{m_L}}+\cdots, \qquad A_0=1,...,N.
\label{tenA0}
\eeq
We see that now string tensions of $N$ $Z_N$ strings are split.

%
%

\section{World sheet effective theory}
 \label{sec:effective_bosonic_action}
\setcounter{equation}{0}

A non-Abelian string has both translational  and orientational zero modes. If we allow a slow dependence of the associated moduli on the world sheet coordinates $z=x_3$ and $t$ they become fields in the effective two dimensional low energy theory on the  string world sheet \cite{ABEKY,SYmon}, see \cite{SYrev} for a review.
Namely, we will have translational moduli $x_{0}^i(t,z)$ (position of the string in the $(x_1,x_2)$ plane, $i=1,2$) and orientational moduli $n^l(t,z)$, $l=1,...,N$. Translational sector is free and decouples therefore we will focus on the orientational sector.

In this section we will derive the bosonic part of the effective world sheet theory on the string.

\subsection{CP$(N-1)$ model on the string world sheet}

First we consider the limit of equal quark masses \eqref{equalmasses}. In this limit color-flavor symmetry \eqref{c+f} is unbroken and the orientational moduli $n^l$ describe the zero modes of the non-Abelian string. Namely, consider a particular $Z_N$ solution \eqref{string-solution} with $n^l=\delta^{lA_0}$, $A_0=1,...,N$. It  breaks the SU$(N)_{C+F}$ group down to SU$(N-1)\times$U(1). Therefore the SU$(N)_{C+F}$
rotation of the $Z_N$ string solution generates the whole family of solutions (non-Abelian string) parametrized by the vector $n^l$ from the moduli space
\beq
\frac{\text{SU}(N)_{C+F}}{\text{SU}(N-1)\times U(1)} = \text{CP}(N-1).
\label{modspace}
\eeq

Since for equal quark masses the SU$(N)_{C+F}$ group is unbroken there is no world sheet potential for orientational moduli $n^l$. To derive the kinetic term we closely follow the general procedure developed in \cite{ABEKY,SYmon} (see \cite{SYrev} for a review) for the \ntwo case.
We substitute  solution \eqref{string-solution} into four-dimensional action \eqref{theory} assuming slow $t$ and $z$ dependence of the orientational moduli $n^l$.

Once the moduli $n^l$ cease to be constant, the gauge field components $A^{{\rm SU}(N)}_0$ and $A^{{\rm SU}(N)}_3$ also become non-vanishing. We use the {\it ansatz} \cite{ABEKY,SYmon,GSY05} 
\begin{equation}
	A^{{\rm SU}(N)}_{k}=-i\,  \big[ \pt_{k} n \,\cdot \bar{n} -n\,\cdot
	\pt_{k} \bar{n} -2n\,
	\cdot \bar{n}(\bar{n}\,\pt_{k} n)
	\big] \,\rho (r)\, , \quad k=0, 3\,,
	\label{An}
\end{equation}
for these components, where we assume a contraction of the color indices inside the parentheses
in the third term.  We also introduced a new profile function $\rho (r)$ which will be determined through a minimization procedure. 

Substituting \eqref{string-solution} and \eqref{An} into \eqref{theory} we get CP$(N-1)$ model
\begin{equation}
	S^{(1+1)}= 2 \beta\,   \int d t\, dz \,  \left\{(\pt_{k}\, \bar{n}\,
	\pt_{k}\, n) + (\bar{n}\,\pt_{k}\, n)^2\right\}\,,
	\label{cp}
\end{equation}
with the coupling constant $\beta$  given by
\begin{equation}
	\beta=\frac{2\pi}{g^2_2}\, I \,,
	\label{betaI}
\end{equation}
where $I$ is a normalization integral determined by the string profile functions 
integrated over $(x_1,x_2)$ plane.
\begin{equation}
	\begin{aligned}
		I & = &	\int_0^{\infty}	rdr\left\{\left(\frac{d}{dr}\rho (r)\right)^2 + \frac{1}{r^2}\, f_{W}^2\,\left(1-\rho \right)^2
		\right.\\[4mm]
		& + &
		\left.  g_2^2\left[\frac{\rho^2}{2}\left(\phi_1^2
		+\phi_2^2\right)+
		\left(1-\rho \right)\left(\phi_2-\phi_1\right)^2\right]\right\}\, .
	\end{aligned}
\label{I}
\end{equation}
The above functional determines an equation of motion for $\rho$,
\begin{equation}
	-\frac{d^2}{dr^2}\, \rho -\frac1r\, \frac{d}{dr}\, \rho
	-\frac{1}{r^2}\, f_{W}^2 \left(1-\rho\right)
	+
	\frac{g^2_2}{2}\left(\phi_1^2+\phi_2^2\right)
	\rho
	-\frac{g_2^2}{2}\left(\phi_1-\phi_2\right)^2=0\, ,
	\label{rhoeq-none}
\end{equation}
while  boundary conditions for $\rho$ read 
\begin{equation}
	\rho (\infty)=0 , \ \ \rho (0)=1 ,
\end{equation}
see \cite{SYrev} for the details.

The formulas above are valid for a non-Abelian string in  both \ntwo QCD and  $\mu$-deformed
QCD. For our large $\mu$  limit we use the string profile functions found in Sec. \ref{sec:profile}. To find the solution for $\rho$ we first note that inside the core
$\rho\approx 1$. In the intermediate region \eqref{logregion} 
$f_W \approx 0$. The terms of equation \eqref{rhoeq-none} which involve derivatives of $\rho$ are negligible compared to the others (we will check that afterwards), and so an approximate solution can be easily found:
\begin{equation}
	\rho ~~\approx~~ \frac{\left(\phi_1-\phi_2\right)^2}{\left(\phi_1^2+\phi_2^2\right)}
		 ~~\approx~~ \frac{\left( \frac{\ln \frac{1}{r m_L}}{\ln \frac{1}{R_g m_L}} \right)^2}%
		 {2 ~-~ 2\frac{\ln \frac{1}{r m_L}}{\ln \frac{1}{R_g m_L}} ~+~ \left( \frac{\ln \frac{1}{r m_L}}{\ln \frac{1}{R_g m_L}} \right)^2},
\end{equation}
where we used solutions \eqref{string:boson:large_mu:dm=0:intermediate} for quark profile functions.

It is easy to check that $\rho' \sim m_L\rho$ and $\rho'' \sim m_L^2\rho$, so it was indeed consistent to drop the derivatives out of the equation \eqref{rhoeq-none}.

Our next step is to substitute this solution into the \eqref{I} and calculate $I$. As we will see, only the region $r \lesssim 1/m_L$ gives a significant contribution to this integral. We have:
\begin{multline}
	I ~~\approx~~ g_2^2 \int r dr \left( \frac{1}{2} \frac{\left(\phi_1-\phi_2\right)^4}{\left(\phi_1^2+\phi_2^2\right)} ~+~
	 \frac{2 \phi_1 \phi_2 \left(\phi_1-\phi_2\right)^2}{\left(\phi_1^2+\phi_2^2\right)} \right) \\
	 ~~=~~ \frac{g_2^2}{2}  \int r dr \frac{\left(\phi_1^2-\phi_2^2\right)^2}{\left(\phi_1^2+\phi_2^2\right)}
\end{multline}
Calculation yields
\begin{equation}
	I ~~\approx~~ c \,\frac{m_G^2}{m_L^2}\,\frac{1}{\ln^2 \frac{m_G^2}{m_L^2}} ~~\sim~~  \frac{g^2|\mu|}{|m|}\,\frac{1}{\ln^2 \frac{g^2|\mu|}{|m|}},
	\label{Iresult}
\end{equation}
where we used Eqs. \eqref{massGSUN} and  \eqref{massLSUN} while the constant $c$ is associated with the ambiguity of the upper limit ($\sim 1/m_L$) of the integral above.
As for the region of large $r$, $r \gg 1/m_L$, the function $\rho$ falls off exponentially, and a contribution from this region is therefore negligible.

Substituting \eqref{Iresult} into \eqref{betaI}  we get the final result for the coupling $\beta$ of the world sheet CP$(N-1)$ model  \eqref{cp}
\begin{equation}
	\beta ~~\approx ~~ c\,\frac{2\pi}{g^2_2}  \,\frac{m_G^2}{m_L^2}\,\frac{1}{\ln^2 \frac{m_G}{m_L}}
	\sim \frac{|\mu|}{|m|}\,\frac{1}{\ln^2 \frac{g^2|\mu|}{|m|}}.
	\label{beta}
\end{equation}

 CP$(N-1)$ model \eqref{cp} is a low  energy effective theory on the string world sheet. It describes
 the  dynamics of massless orientational moduli at energies much below the inverse thickness of the string proportional to $m_L$. If we go to higher energies we have to take into account higher derivative 
 corrections to \eqref{cp}. 

Relation \eqref{beta} is derived at the classical level. In quantum theory 
the coupling constant $\beta$ runs. Relation \eqref{beta} defines the CP$(N-1)$ model coupling at
 a scale of the ultra-violet (UV) cutoff of the world sheet theory equal to $m_G$.
 In fact,  CP$(N-1)$ model is an asymptotically free theory. Its coupling at the UV scale 
$m_G\sim \sqrt{\xi}$
 at one loop is given by 
\beq
4\pi \beta(\sqrt{\xi})= N\ln{\frac{\sqrt{\xi}}{\Lambda_{CP}}},
\eeq
where  $\Lambda_{CP}$ is the scale of the CP$(N-1)$ model. This gives for the scale $\Lambda_{CP}$
\beq
\Lambda_{CP} \approx \sqrt{\xi} \exp{- {\rm const}\,\frac{|\mu|}{|m|}\frac{1}{\ln^2 \frac{g^2|\mu|}{|m|}}}.
\label{LambdaCP}
\eeq

We see that the scale of CP$(N-1)$ model $\Lambda_{CP}$ is exponentially small, so the world sheet theory
is weekly coupled in a wide region of energies $\gg \Lambda_{CP}$. This should be  contrasted to non-Abelian string in \ntwo QCD where world sheet theory has a scale $\Lambda_{CP}$ equal to scale 
$\Lambda_{{\cal N}=2}$ of
the bulk QCD \cite{SYrev}.

\subsection{World-sheet potential  at large $\mu$}
\label{sec:worldsheetpot}

In this subsection we relax the condition of equal quark masses \eqref{equalmasses} and consider the effect
of quark mass differences to the leading order in $\Delta m_{AB}$, see \eqref{smallmassdiff}.
Non-equal quark masses break color-flavor symmetry \eqref{c+f} down to U$(1)^N$ so 
as we already mentioned above the orientational modes of the non-Abelian string are no longer zero modes. They become quasizero
modes in the approximation of small quark mass differences \eqref{smallmassdiff}. We still can introduce the orientational quasimoduli $n^l$, $l=1,...,N$ and consider a shallow  potential in the CP$(N-1)$ world sheet theory \eqref{cp} generated by the mass differences.
 We neglect effects of small mass differences in the kinetic term assuming that it is still given by 
 Eq. \eqref{cp}.
 
Our general strategy is to take string solution \eqref{string-solution} with the unperturbed string profile functions of Sec. \ref{sec:profile} and substitute it into potential \eqref{potential:3+1:F:original}
 taking into account explicit $m_A$ dependence of this potential to the leading order in $\Delta m_{AB}$.
 After a rather involved  calculation we arrive at the potential of the world sheet theory
 \begin{equation}
 	\delta V_{1+1} ~~=~~ \chi   \sum_{A=1}^{N} \frac{\Re \left[(\xi_A - \hat{\xi}) \bar{\hat{\xi}}\right]}{|\hat{\xi}|} |n^A|^2 ,
 	\label{V1+1-general}
 \end{equation}
 where $\delta V_{1+1}$  is the potential up to a constant, $\xi_P$ are given by \eqref{xi-general}, while the factor $\chi$ is determined  by the
string profile functions integrated over $(x_1,x_2)$ plane,
\begin{equation}
	\chi ~~=~~ \frac{\pi }{|\mu_2|^2} \int\limits_{0}^{\infty} rdr\, (\phi_2^2 - \phi_1^2)\,
	\left[\phi_1^2 - \frac{\alpha}{N}(\phi_1^2 - \phi_2^2)\right].
	\label{chi}
\end{equation}

Now we use our solutions for $\phi_1$ and $\phi_2$ (see Sect. \ref{sec:profile}) and integrate here over the region $r\lsim 1/m_L$. We also  assume that
$\mu_1$ and $\mu_2$ scale in such a way that the parameter $\alpha$ in \eqref{alpha} is fixed. More explicitly,	 we assume that
\begin{equation}
\mu ~~\equiv~~\mu_2 ~~=~~ {\rm const} \cdot \mu_1 \sqrt{\frac{2}{N}}  ,
\label{mus}
\end{equation} 
This gives for $\chi$
\begin{equation}
	\chi ~~\approx~~ {\rm const} \cdot \frac{2\pi}{\ln \frac{m_G }{m_L}} .
\end{equation}
 Moreover, the region of integration $r\gg 1/m_L$ in \eqref{chi} does not contribute to the leading order. The unknown constant
 above appears due to an ambiguity of the upper limit of the integral over $r$, $r\sim 1/m_L$.

Substituting this into \eqref{V1+1-general} we get
\begin{equation}
	\delta V_{1+1} ~~\approx~~ {\rm const} \cdot 2\pi   \sum_{A=1}^{N}
			\frac{ \Re \left[(\xi_A - \hat{\xi}) \bar{\hat{\xi}}\right]}{|\hat{\xi}|
			\ln \frac{m_G }{m_L}} |n^A|^2 .
			\label{deltaV1+1}
\end{equation} 

Now let us  fix the unknown constant in the equation above comparing it with expressions \eqref{tenA0} for the string tensions of $Z_N$ strings. $Z_N$ strings are extreme points of the world sheet potential $V_{1+1}$, therefore a 
value of $V_{1+1}$ at the extreme point $n^l = \delta^{lA_0}$ corresponding to $A_0$-th $Z_N$ 
string should be equal
to string tension \eqref{tenA0}.
This gives ${\rm const} =1$ in \eqref{deltaV1+1} and leads us to the final expression for the potential in the world sheet CP$(N-1)$ theory 
\begin{equation}
		V_{1+1} ~~\approx~~ \frac{4\pi}{\ln \frac{ m_G}{m_L}}  \,  \abs{\sqrt{\frac{2}{N}} \mu_1 \hat{m} ~+~ \mu_2 \left( \sum\limits_{A=1}^{N} m_A |n^A|^2 - \hat{m} \right) }.
					\label{V1+1}
\end{equation} 
This potential integrated over world sheet coordinates $t$ and $z$ should be added to the kinetic term 
in \eqref{cp}.
Note, that this potential is a generalization of our result in \eqref{deltaV1+1} since it includes all terms in the expansion in powers of $(m_A-\hat{m})/\hat{m}$. For the single trace $\mu$-deformation \eqref{mu_condition}
the world sheet potential takes a particularly simple form
\begin{equation}
		V_{1+1} ~~\approx~~ \frac{4\pi}{\ln \frac{ m_G}{m_L}} \,   |\mu_2|\left| \sum\limits_{A=1}^{N} m_A |n^A|^2  \right| .
					\label{V1+1singletr}
\end{equation} 

The potential \eqref{V1+1} has only one minimum and one maximum at generic $\Delta m_{AB}$. Other $(N-2)$ extreme points are saddle points. All these extreme points  are located at
\beq
n^A = \delta^{AA_0}, \qquad A_0=1,...,N,
\label{saddleloc}
\eeq
and associated with the $Z_N$ strings. 
A value of the potential at a given extreme point coincides with the tension of the $A_0$-th $Z_N$ string,
\beq
V_{1+1}(n^A = \delta^{AA_0}) = T_{A_0}, \qquad A_0=1,...,N.
\label{minima}
\eeq
Absolute minimum (the unique vacuum)  of \eqref{V1+1} corresponds to the $Z_N$ string associated with winding of a squark with the smallest mass.

Note that our derivation of Eq. \eqref{deltaV1+1} reproduced the logarithmic suppression typical
for string tensions in extreme type I superconductors (with small Higgs mass $m_L$), see \eqref{tenA0}
and \cite{Y99}.

Potential \eqref{V1+1} is similar to the potential in the world sheet theory on the non-Abelian string
derived in \cite{SYfstr} for $\mu$-deformed \ntwo QCD in the limit of  small $\mu$. In this case the world sheet
theory is heterotic CP$(N-1)$ model with \ntwoo supersymmetry. For small $\mu$ the world sheet potential
obtained in \cite{SYfstr} can be written in the form
\beq
	V_{1+1}^{\mu\to 0} ~~=~~ 4\pi     \abs{\sqrt{\frac{2}{N}} \mu_1 \hat{m} ~+~ \mu_2 \left( \sum\limits_{A=1}^{N} m_A |n^A|^2 - \hat{m} \right) }.
					\label{V1+1SYfstr}
\eeq
It differs from the one in \eqref{V1+1} by the absence of the logarithmic suppression. This has a natural explanation. At small $\mu$ in much the same way as in our case the saddle points of the world sheet potential correspond to the $Z_N$ strings and relation \eqref{minima} is still valid.
On the other hand in the limit of small $\mu$ the $Z_N$ strings are BPS saturated and their tensions are given by
$T_{A_0}^{\mu\to 0} = 2\pi|\xi_{A_0}|$, see \cite{SYfstr}. This explains the absence of the logarithmic suppression in the potential \eqref{V1+1SYfstr}.

\subsection{Mass spectrum on the string}

Let us assume that $m_1$ is the smallest quark mass. Then the vacuum of the world sheet potential \eqref{V1+1} is located at 
\beq
n^A = \delta^{A 1}
\label{minimumloc}
\eeq
and the minimum value $V_{1+1}^{\rm min} = T_{A=1}$. Let us calculate a perturbative mass spectrum of 
the world sheet theory in this vacuum. Expanding 
\begin{equation}
	|n^{1}|^2 ~~=~~ 1 ~-~ \sum_{A\neq 1} |n^A|^2
\end{equation}
and extracting the quadratic in fluctuations $n^A$ terms from potential \eqref{V1+1}, 
we get masses of world sheet excitations $n^A$, $A\neq 1$
\begin{equation}
	m_{A \neq 1}^2 ~~=~~ \frac{\pi}{\beta \ln \frac{ m_G}{m_L}} 
	\frac{ \Re \left[(\xi_A - \xi_{1}) \bar{\xi_{1}}\right]}{|\xi_{1}|} .
	\label{masses2Dgen}
\end{equation}
Note the factor $1/(2\beta)$ here that comes from the kinetic term in \eqref{cp}. Substituting here the coupling $\beta$ from \eqref{beta} we see that the masses of the perturbative world sheet excitations behave as
\beq
m_{A \neq 1}^2 \sim m (m_A-m_1)\,\ln \frac{ m_G}{m_L}
\label{masses2D}
\eeq
 
The coupling constant of CP$(N-1)$ model grows at low energies and gets frozen at the scale of 
the masses calculated above. If these masses are much larger than $\Lambda_{CP}$ \eqref{LambdaCP} then the world sheet theory is at weak coupling. Since $\Lambda_{CP}$ is exponentially small we see that world sheet theory is 
at weak coupling even at rather small mass differences $\Delta m_{AB}$. However in the equal quark mass limit
\eqref{equalmasses} when $\Delta m_{AB} =0$ the world sheet CP$(N-1)$ model becomes strongly coupled.

Our result \eqref{V1+1} for the world sheet potential on the non-Abelian string in $\mu$-deformed theory can be  compared with the world sheet potential for the non-Abelian string in \ntwo supersymmetric QCD with FI $D$-term
generated by quark mass differences, see \cite{SYrev} for a review. In the \ntwo case all the $Z_N$ strings are degenerate, with tensions given by the FI parameter. The world sheet potential in this case has $N$  minima
located at \eqref{saddleloc} separated by shallow barriers quadratic in $\Delta m_{AB}$. The world sheet
theory has \ntwot supersymmetry and the presence of $N$ vacua is ensured by the Witten index for CP$(N-1)$ 
supersymmetric model. There are kinks interpolating between these vacua which are interpreted as confined monopoles of bulk QCD \cite{SYmon,HT2}, see Sec.~\ref{physics} and \cite{SYrev} for a review.

In the limit of large $\mu$ potential \eqref{V1+1} dominates over the quadratic in $\Delta m_{AB}$
potential, and one can neglect the latter one. We see that most of the vacua present in the \ntwo case are lifted and
the world sheet theory has a single vacuum at non-zero  $\Delta m_{AB}$. Moreover, the lifted vacua are 
saddle points rather than  local minima and therefore classically they are unstable. This means that there are no kinks 
in the world sheet theory. 

Thus we come to the  conclusion that confined monopoles present in \ntwo QCD with FI term do not survive
large $\mu$ limit when $\mu$-deformed theory flows to \none QCD, provided that $\Delta m_{AB}$ are non-zero.
Only when $\Delta m_{AB}=0$, the potential \eqref{V1+1} vanishes (and the world sheet theory enters into the strong
coupling), and we can consider kinks/confined monopoles. We will discuss this case below in Sec.~\ref{physics}

%
%

\section{Fermion zero modes
\label{sec:superorient-zero}}
\setcounter{equation}{0}

In this section we consider the fermion zero modes of the non-Abelian string. First we briefly review  
the limit of 
small $\mu$, see \cite{SYrev} for a more detailed review. In this limit deformation superpotential 
\eqref{none_superpotential_general} reduces to the FI $F$-term and does not break \ntwo supersymmetry
\cite{HSZ,VY}. In the \ntwo limit both superorientational and supertranslational fermion zero modes 
of the non-Abelian string can be obtained by a supersymmetry transformation of the bosonic string solution
\cite{SYmon,SY02,BSYhet}. Next we gradually increase $\mu$ and study perturbations of 
superorientational zero modes at small $\mu$. We   show that all the superorientational fermion zero 
modes are lifted by the $\mu$-deformation. As a result fermionic moduli which become fermion fields in the 
two dimensional low energy $CP(N-1)$ model on the string acquire masses. Eventually they disappear from
the world sheet theory in the large $\mu$ limit. Finally we  comment on supertranslational
fermion zero modes which in much the same way as in \ntwo theory can be obtained by supersymmetry 
transformations from the bosonic string solution.

\subsection{Superorientational modes in \ntwo limit}

The fermionic part of the \ntwo QCD defined by superpotentials \eqref{supN2} and 
\eqref{none_superpotential_general} (before integrating out adjoint fields)  is as follows:
\begin{align}
\notag
\mc{L}_{\rm 4d} & ~~=~~ \frac{2i}{g_2^2} \Tr\, \ov{\lN_f \slashed{\md}} \lambda^{f\text{SU($N$)}}
~+~ \frac{i}{g_1^2} \ov{\lU_f \slashed{\p}} \lambda^{f{\rm U(1)}}
~+~ \Tr\, i\, \ov{\psi \slashed{\nabla}} \psi
~+~ \Tr\, i\, \wt{\psi} \slashed{\nabla} \ov{\wt{\psi}}
\\[3mm]
\notag
&
~+~
i\sqrt{2}\, \Tr \lgr \ov{q}{}_f \lambda^{f{\rm U(1)}}\psi
~+~ \wt{\psi} \lU_f q^f
~+~ \ov{\psi \lU_f} q^f
~+~ \ov{q^f \lU_f \wt{\psi}}
\rgr
\\[3mm]
\label{fermact}
&
~+~
i\sqrt{2}\, \Tr \lgr \ov{q}{}_f \lambda^{f\text{SU($N$)}} \psi
~+~ \wt{\psi} \lN_f q^f
~+~ \ov{\psi \lN_f} q^f
~+~ \ov{q^f \lN_f \wt{\psi}}
\rgr
\\[3mm]
\notag
&
~+~
i\sqrt{2}\, \Tr\; \wt{\psi} \left( \frac12 \aU ~+~ \frac{m_A}{\sqrt{2}} ~+~ \aN \right) \psi
~+~
i\sqrt{2}\, \Tr\; \ov{\psi} \left( \frac12\baU ~+~ \frac{m_A}{\sqrt{2}} ~+~ \baN \right) \ov{\wt{\psi}}
\\[3mm]
\notag
&
~-~
2 \sqrt{\frac{N}{2}} \mu_1 \lgr \left( \lambda^{2\,{\rm U(1)}} \right)^2
~+~ \left( \ov{\lambda}{}^{\rm U(1)}_2 \right)^2 \rgr
~-~
\mu_2 \Tr \lgr \left( \lambda^{2\,\text{SU($N$)}} \right)^2
~+~ \left( \ov{\lambda}{}^\text{SU($N$)}_2 \right)^2 \rgr\,,
\end{align}
where derivatives acting on fermion fields are defined by the $\sigma$-matrices, for example
  $\bar{\slashed{\nabla}}= \nabla_{\mu}\overline{\sigma}{}^\mu_{\dot{\alpha}\alpha}$, and  
a color-flavor matrix notation is used for the quark fermions $\psi_{\alpha}^{kA}$, $\tilde{\psi}^{\alpha}_{Ak}$. 
Index $f$ 
is SU(2)$_R$ index of the \ntwo theory,  $ q^f ~=~ (q, \ov{\wt{q}}) $, 
$\lambda_{\alpha}^f=(\lambda_{\alpha}^1,\lambda_{\alpha}^2)$. Note the 
$\mu$-deformation mass terms for the $f = 2$ gauginos in \eqref{fermact}. In the \ntwo limit these terms 
vanish.

A string solution in the \ntwo limit at small $\mu$ is 1/2 BPS, which means that half of the supercharges 
of the \ntwo theory act trivially on solution \eqref{string-solution}, provided the orientational vector
$n^l$ is a constant vector. Namely, the four supercharges (out of eight supercharges $Q^{\alpha f}$)
 that satisfy the constraints 
\beq
Q^{21} ~=~ Q^{22}, \qquad\qquad   Q^{11} ~=~ - Q^{12} \ .
\label{trivQ}
\eeq
act trivially on the BPS string in the \ntwo theory with the FI $F$-term \cite{VY,SYmon,SYrev}.

The other four supercharges  generate 
four supertranslational modes which are superpartners of the two translational modes.

However once the orientational vector
$n^l$ acquires a slow $t$ and $z$ dependence, the supercharges selected by \eqref{trivQ} 
become supersymmetry generators acting in the \ntwot supersymmetric $CP(N-1)$ model on the 
string world sheet \cite{SYmon}. This allows one to obtain the orientational fermionic zero modes
from a bosonic solution using supersymmetry transformations selected by \eqref{trivQ} \cite{SYmon,SYrev}.
The result is 
\begin{align}
\notag
\ov{\psi}{}_{\dot{2}} & ~~=~~  \phantom{-\, }
\frac{\phi_1^2 ~-~ \phi_2^2}{\phi_2}\cdot n \bar{\xi}_L~,
\\[2mm]
\notag
\ov{\wt{\psi}}{}_{\dot{1}} & ~~=~~   -\,
\frac{\phi_1^2 ~-~ \phi_2^2}{\phi_2}\cdot \xi_R\bar{n} ~,
\\[2mm]
\label{N2_sorient}
\lambda^{11\ \text{SU($N$)}}  & ~~=~~ \phantom{-\, } i \frac{\phi_1}{\phi_2}\, f_W \, \frac{\displaystyle x^1 - i\, x^2}{\displaystyle r^2}\cdot
n \bar{\xi}_L
\\[2mm]	
\notag
\lambda^{22\ \text{SU($N$)}}  & ~~=~~ -\,  i \frac{\phi_1}{\phi_2}\, f_W \, \frac{\displaystyle x^1 + i\, x^2}{\displaystyle r^2}\cdot
\xi_R\bar{n}
\\[2mm]
\notag
\lambda^{12\ \text{SU($N$)}}  & ~~=~~ \lambda^{11\ \text{SU($N$)}}
\, ,\qquad    \lambda^{21\ \text{SU($N$)}}   ~~=~~ -\, \lambda^{22\ \text{SU($N$)}}\,,
\end{align}	
where we suppress color and flavor indices while superscripts of adjoint fermions mean $\lambda^{\alpha f}$.

Note that the bosonic profile functions of the string $\phi_{1,2}(r)$, $f(r)$ and $f_W$  in this 
section are the profile 
functions of the BPS string in the \ntwo limit of small $\mu$ rather than the string profile 
functions of 
Sec. \ref{sec:profile}, which corresponds to the large $\mu$ limit. Former are solutions of first order 
equations rather than second order equations \eqref{profileeqs}. They satisfy boundary conditions 
\eqref{boundary} and were found numerically in \cite{ABEKY}, see \cite{SYrev} for a review.

The Grassmann variables $ \xi_{R,L}^l$, $l=1,...,N$ in \eqref{N2_sorient} are proportional to  the parameters of 
the supersymmetry transformations $\epsilon^{\alpha f}$ selected by \eqref{trivQ}, namely
\beq
 \xi_L^l \sim \epsilon^{21}+\epsilon^{22}, \qquad \xi_R^l \sim \epsilon^{12}-\epsilon^{11}.
\label{xi's}
\eeq
These parameters become fermion fields (superpartners of $n^l$) in the effective world sheet $CP(N-1)$
 model once we allow their slow 
dependence on the world sheet coordinates $t$ and $z$ \cite{SYmon,SYrev}.
They are subject to the conditions
\begin{equation}
 \label{constr}
    \nbar_l\, \xi^l_{L,R}   ~~=~~ 0~,
\end{equation}
which are a supersymmetric generalization of the CP($N-1$) condition $ |n|^2 = 1 $.

\subsection{Small $\mu$ expansion for fermion orientational zero modes}

As we switch on the mass terms for the $f=2$ gauginos (see the last line in \eqref{fermact}), 
the theory becomes \none supersymmetric and half of the supercharges $Q^{\alpha f=2}$ are lost. 
There are no SUSY 
transformations which act trivially on the string with constant $n^l$ (they were used to 
generate superorientational modes in \ntwo limit), and the string is no longer BPS. Therefore, to 
calculate zero modes one has to solve the Dirac equations. 

Note that for the case of the massless $\mu$-deformed theory with FI $D$-term considered in \cite{BSYhet}
the supercharges that act trivially on the string with constant $n^l$ in the \ntwo limit
 are $Q^{12}$ and $Q^{21}$ instead of the linear combinations selected by \eqref{trivQ}. 
 Therefore as we switch on the $\mu$-deformation only one (two real) of the above supercharges 
is lost, namely $Q^{12}$. The other one (two real), $Q^{21}$, still 
acts trivially and ensures that the string is 
still BPS-saturated. In our case all four supercharges of \none theory $Q^{\alpha 1}$ act non-trivially
on the string. This is the reason why the string ceases to be a BPS one as we switch on $\mu$.

Dirac equations which follow from action \eqref{fermact} read
\begin{align}
\notag
&\phantom{-i}
\frac{i}{g_1^2}\, \left( \ov{\slashed{\p}}\lambda^{f {\rm U(1)}} \right)
~+~  i \sqrt{2}\, \Tr\lgr \ov{\psi} q^f  ~+~ \ov{q}{}^f \ov{\wt{\psi}} \rgr
~-~ 4\, \delta_2^{\ f}\sqrt{\frac{N}{2}} \mu_1\, \ov{\lambda}^{U(1)}_2  ~~=~~ 0\,,
\\[2mm]
\notag
&\phantom{-i}
\frac{i}{g_2^2}\, \left( \ov{\slashed{\md}}\lambda^{f \text{SU($N$)}}\right)^a
~+~ i \sqrt{2}\, \Tr\lgr \ov{\psi}T^a q^f  ~+~  \ov{q}{}^f T^a \ov{\wt{\psi}} \rgr
~-~ \delta_2^{\ f} \mu_2\, \ov{\lambda}{}_2^{a \text{SU($N$)}}  ~~=~~ 0 \,,\\[2mm]
\notag
&
-i\, \ov{\psi} \overleftarrow{\ov{\slashed{\nabla}}}
~+~ i \sqrt{2} \lgr \ov{q}{}_f \left\{ \lambda^{f {\rm U(1)}} + \lambda^{f \text{SU($N$)}} \right\}
~+~ \wt{\psi} \left\{ \frac12\baU ~+~ \frac{m_A}{\sqrt{2}} ~+~ \baN \right\} \rgr    ~~=~~ 0\,, \\[2mm]
\label{dirac}
&\phantom{-i}
i\, \slashed{\nabla} \ov{\wt{\psi}}
~+~ i \sqrt{2} \lgr \left\{ \lambda_f^{\rm U(1)} ~+~ \lambda_f^\text{SU($N$)} \right\} q^f
~+~ \left\{ \frac12\baU ~+~ \frac{m_A}{\sqrt{2}} ~+~ \baN \right\}\psi \rgr  ~~=~~ 0\,, \\[2mm]
\notag
&\phantom{-i}
i \ov{\slashed{\nabla}}\psi
~+~ i \sqrt{2} \lgr \left\{ \ov{\lambda}{}^{\rm U(1)}_f ~+~ \ov{\lambda}{}^\text{SU($N$)}_f \right\} q^f
~+~ \left\{\frac12 \aU ~+~ \frac{m_A}{\sqrt{2}} ~+~ \aN \right\} \ov{\wt{\psi}} \rgr   ~~=~~ 0 \,,\\[2mm]
\notag
&
-i\, \wt{\psi} \overleftarrow{\slashed{\nabla}}
~+~ i \sqrt{2} \lgr \ov{q}{}^f \left\{ \ov{\lambda}{}_f^{\rm U(1)} ~+~ \ov{\lambda}{}_f^\text{SU($N$)} \right\}
~+~ \ov{\psi} \left\{ \frac12\aU ~+~ \frac{m_A}{\sqrt{2}} ~+~ \aN \right\} \rgr  ~~=~~ 0\,.
\end{align}

To simplify the problem we will use below the following strategy. We consider the region of small $\mu$
and look for solutions of the Dirac equations above  perturbatively in $\mu$.
Of course,  zero modes \eqref{N2_sorient}  satisfy  Dirac equations \eqref{dirac} \cite{SY02,BSYhet}
at $\mu=0$. 
We take these modes as a zero order solutions and solve for perturbations proportional to $\mu$.

Similar method was used in \cite{SY02,BSYhet} for a massless $\mu$-deformed theory with a FI $D$-term.
In that case it was  shown that the orientational fermion zero modes survive the $\mu$-deformation,
however, their profile functions become deformed. Below we will show that in our case of $\mu$-deformed 
theory  with massive quarks without the FI $D$-term the answer is different: orientational fermion 
zero modes do not survive the $\mu$-deformation.

In analogy with the method of \cite{BSYhet} we will use an {\it ansatz} for 
the superorientational modes:
\begin{equation}
 \begin{aligned}
  \lambda^{1f\ \text{SU($N$)}} & ~~=~~ 2\, \frac{x^1 - ix^2}{r}\, \lambda^{1f}_+(r) \; n\ov{\xi}_L
  ~~+~~  2\, \lambda^{1f}_-(r)\; \xi_L\ov{n}\,,
  \\[2mm]
  \lambda^{2f\ \text{SU($N$)}} & ~~=~~ 2\, \frac{x^1 + ix^2}{r}\, \lambda^{2f}_+(r) \; \xi_R\ov{n}
  ~~+~~  2\, \lambda^{2f}_-(r)\; n\ov{\xi}{}_R\,,
 \end{aligned}
 \label{small_mu-supertransl-ansatz-lambda}
\end{equation}
\begin{equation}
\begin{aligned}
  \ov{\wt{\psi}}{}_{\dot{1}} & ~~=~~ 2\, \ov{\wt{\psi}}_{\dot{1}+}(r)\;  \xi_R \ov{n}
  ~~+~~  2\, \frac{x^1 - i x^2}{r}\, \ov{\wt{\psi}}_{\dot{1}-}(r)\; n\ov{\xi}{}_R~.
  \\[2mm]
  \ov{\wt{\psi}}{}_{\dot{2}} & ~~=~~ 2\, \ov{\wt{\psi}}_{\dot{2}+}(r)\; n\ov{\xi}_L
  ~~+~~  2\, \frac{x^1 + i x^2}{r}\, \ov{\wt{\psi}}_{\dot{2}-}(r)\; \xi_L\ov{n}~.
  \\[2mm]
  \ov{\psi}{}_{\dot{1}} & ~~=~~ 2\, \ov{\psi}_{\dot{1}+}(r)\;  \xi_R \ov{n}
  ~~+~~  2\, \frac{x^1 - i x^2}{r}\, \ov{\psi}_{\dot{1}-}(r)\; n\ov{\xi}{}_R~.
  \\[2mm]
  \ov{\psi}{}_{\dot{2}} & ~~=~~ 2\, \ov{\psi}_{\dot{2}+}(r)\; n\ov{\xi}_L
  ~~+~~  2\, \frac{x^1 + i x^2}{r}\, \ov{\psi}_{\dot{2}-}(r)\; \xi_L\ov{n}~.
\end{aligned}
\label{small_mu-supertransl-ansatz-psi}
\end{equation}
Here $ \lambda_+(r) $ and $ \psi_+(r) $ represent "undeformed" profile functions  present 
in the \ntwo case, while $ \lambda_-(r) $ and $ \psi_-(r) $ are the "perturbations" due to $\mu$-deformation.
 Of course this terminology makes sense only in the small $\mu$ limit, then "-" -components will 
be of order $\mu$. More generally, the ''+'' profile functions are expanded in even powers of $\mu$,
while ''-'' components are expanded in odd powers of $\mu$.

Let us consider the equations for the perturbative "-" -components (solutions for the "+" -components 
are given   by \eqref{N2_sorient} up to the $O(\mu^2)$ terms). Half of them are very similar to those 
solved in \cite{BSYhet}. If we denote
\[
\lambda^{22}_- - \lambda^{21}_- \equiv \lambda_-\, ,
\]
then two of these equations for $\lambda_{-} $ and $\ov{\wt{\psi}}_{\dot{1}-}$ take the form
\beqn
&&
\p_r \ov{\wt{\psi}}_{\dot{1}-}(r) ~+~ \frac{1}{r}\ov{\wt{\psi}}_{\dot{1}-}(r) ~-~ 
\frac{1}{Nr}\left(f+f_W(N-1)\right) \, \ov{\wt{\psi}}_{\dot{1}-}(r) ~+~ i\phi_2\,\lambda_-~~=~~ 0
\nonumber \\
&&
-\,  \p_r\lambda_- ~-~ \frac{f_W}{r}\lambda_-
~+~ i\,g_2^2 \phi_2\, \ov{\wt{\psi}}_{\dot{1}-}(r) ~-~ \mu_2 g_2^2\, \frac{i}{2} \frac{f_W}{r} \frac{\phi_1}{\phi_2} ~~=~~ 0
\label{easy}
\eeqn
These equations can be solved in much the same way as in \cite{BSYhet}. The solutions are
\begin{align}
\notag
\ov{\wt{\psi}}_{\dot{1}-} & ~~=~~ -\, \mu_2 g_2^2 \,\frac{r}{8\phi_1} \left( \phi_1^2 ~-~ \phi_2^2 \right)  
~+~ O(\mu^3)\,,
\\[2mm]
\label{firstorder-ovwt1}
\lambda_- & ~~=~~ \lambda^{22}_- - \lambda^{21}_- ~~=~~ -\, \mu_2 g_2^2 \,\frac{i}{4} \lgr (f_W - 1) \frac{\phi_2}{\phi_1} ~+~ \frac{\phi_1}{\phi_2} \rgr ~+~ O(\mu^3)~.
\end{align}
Another pair of the profile functions $\ov{\psi}_{\dot{2}-}$ and $(\lambda^{12}_- ~+~ \lambda^{11}_-)$
satisfies the same equations \eqref{easy}.  Hence the solution reads
\begin{align}
\notag
\ov{\psi}_{\dot{2}-} & ~~=~~ \ov{\wt{\psi}}_{\dot{1}-}\, , 
\\[2mm]
\label{firstorder-ov2}
\lambda^{12}_- ~+~ \lambda^{11}_- &~~=~~ \lambda_- .
 \end{align}

Let us study behavior of these solutions in the limits $r\to\infty$ and $r\to 0$.
The bosonic  profile functions fall off exponentially at infinity
\begin{equation}
 f_W(r) \sim \exp{-m_G r}, \qquad \phi_{1,2}-\sqrt{\xi}\sim \exp{-m_G r},
\label{infbehb}
\end{equation}
while their behavior at $r\to 0$ is as follows:
\beq
 f_W(r)-1 \sim r^2, \qquad \phi_{1}\sim r, \qquad \phi_2 \sim {\rm const},
\label{zerobehb}
\eeq
see \eqref{boundary}.

From this behavior we see that fermion zero modes \eqref{firstorder-ovwt1}, \eqref{firstorder-ov2}
are normalizable. They fall off exponentially at $r\to\infty$ and are regular at $r\to 0$.

Now consider solutions  for the other components. They turn out to be more complicated. Denoting
\[
\lambda^{22}_- + \lambda^{21}_- \equiv \lambda_{(1)}\, ,
\]
one gets:
\begin{equation}
\begin{aligned}
\p_r \ov{\psi}_{\dot{1}-}(r) ~+~ \frac{1}{r}\ov{\psi}_{\dot{1}-}(r) ~+~ \frac{1}{Nr} (f-f_W) \, 
\ov{\psi}_{\dot{1}-}(r) ~+~ i\phi_1\,\lambda_{(1)} &~~=~~ 0\,,\\[2mm]
-\,  \p_r\lambda_{(1)} ~-~ \frac{f_W}{r}\lambda_{(1)}
~+~ i\,g_2^2 \phi_1\, \ov{\psi}_{\dot{1}-}(r) ~-~ \mu_2 g_2^2\, 
\frac{i}{2} \frac{f_W}{r} \frac{\phi_1}{\phi_2} &~~=~~ 0.
\end{aligned}
\label{fermeqs_ovpsi1_lplus_small-1}
\end{equation}
 So far solutions for our equations were given by certain algebraic combinations of the bosonic profile functions. 
However, for the functions $\ov{\psi}_{\dot{1}-}$ and $\lambda_{(1)}$ it is not the case. 
The above equations are solved in Appendix A. The solutions are given by
Eqs.~ \eqref{fermsol_ovpsi1_lplus_small} and \eqref{fermsol_lamplus_lplus_small}.

Two remaining modes   $\ov{\wt{\psi}}_{\dot{2}-}$ and $(\lambda^{12}_- ~-~ \lambda^{11}_-)$
satisfy the same equations \eqref{fermeqs_ovpsi1_lplus_small-1}.  Therefore  these modes are 
given by the same expressions,


\beq
\ov{\wt{\psi}}_{\dot{2}-}(r)=\ov{\psi}_{\dot{1}-}(r), \qquad 
\lambda^{12}_{-} ~-~ \lambda^{11}_{-}= \lambda_{(1)}(r).
\label{remainingferm}
\eeq

Solutions \eqref{fermsol_ovpsi1_lplus_small} and \eqref{fermsol_lamplus_lplus_small} 
fall off exponentially at infinity, however, the behavior of the field $\lambda$ 
in  \eqref{fermsol_lamplus_lplus_small}  is singular at $r\to 0$, namely
it is proportional to $1/r$.
   This means that these modes are non-renormalizable. Our perturbative approach does not work: 
the corrections
to \eqref{N2_sorient} proportional to $\mu$ turn out to be non-normalizable. We will show in the next 
subsection that
the resolution of this puzzle is that the fermion orientational modes get lifted by the $\mu$-deformation.

 \subsection{Lifted fermion orientational modes}

Let us consider instead of Dirac equations \eqref{dirac}  equations with a non-zero eigenvalue for 
quark fermions, namely
\begin{multline}
  -i\, \ov{\psi} \overleftarrow{\ov{\slashed{\nabla}}}
 ~+~ i \sqrt{2} \Bigg( \ov{q}{}_f \left\{ \lambda^{f {\rm U(1)}} + \lambda^{f \text{SU($N$)}} \right\} \\
 ~+~ \wt{\psi} \left\{ \frac12 \baU ~+~ \frac{m_A}{\sqrt{2}} ~+~ \baN \right\} \Bigg)   
 ~~=~~ - m_{or } \, \wt{\psi} \,,
 \label{dirac_modifyed-ovpsi}
\end{multline}
\begin{multline}
  i\, \slashed{\nabla} \ov{\wt{\psi}}
 ~+~ i \sqrt{2} \Bigg( \left\{ \lambda_f^{\rm U(1)} ~+~ \lambda_f^\text{SU($N$)} \right\} q^f \\
 ~+~ \left\{ \frac12\baU ~+~ \frac{m_A}{\sqrt{2}} ~+~ \baN \right\}\psi \Bigg)  ~~=~~ - m_{or } \,\psi \, .
 \label{app:dirac_modifyed-ovwtpsi}
\end{multline}
with the mass $m_{or}$  to be determined from the condition of normalizability 
of superorientational modes.

Proceeding exactly as it was done in the previous subsection, instead of 
Eqs.~\eqref{fermeqs_ovpsi1_lplus_small-1}
we arrive  at
\begin{equation}
\begin{aligned}
\p_r \ov{\psi}_{\dot{1}-}(r) ~+~ \frac{1}{r}\ov{\psi}_{\dot{1}-}(r) ~+~ \frac{1}{Nr} (f-f_N) \, \ov{\psi}_{\dot{1}-}(r) ~+~ i\phi_1\,\lambda_{(1)} &~~=~~ m_{or }\,\frac{\phi_1^2 ~-~ \phi_2^2}{2\phi_2}\,,\\[2mm]
-\,  \p_r\lambda_{(1)} ~-~ \frac{f_N}{r}\lambda_{(1)}
~+~ i\,g_2^2 \phi_1\, \ov{\psi}_{\dot{1}-}(r) ~-~ \mu_2 g_2^2\, \frac{i}{2} \frac{f_N}{r} \frac{\phi_1}{\phi_2} &~~=~~ 0.
\end{aligned}
\label{fermeqs_ovpsi1_lplus_small-2}
\end{equation}
We consider these equations in the  Appendix A. The solutions are given by  \eqref{solution_orient_ov_psi_1-}
and \eqref{solution_orient_lambda(1)}.
The condition of regularity of these solutions at $r\to 0$ gives the eigenvalue 
\begin{equation}
m_{or } ~~=~~ -\frac{\displaystyle{\mu_2 g_2^2}\,\int\limits_0^\infty {\mathrm d}y \frac{f_N^2(y)\phi_1^2(y)}{y\phi_2^2(y)}}{ 1 ~-~ 2\displaystyle\int\limits_0^\infty {\mathrm d}y \frac{f_N^2(y)\phi_1^2(y)}{y\phi_2^2(y)}}.
\label{solution_orient_alpha}
\end{equation}
Solutions for $\ov{\wt{\psi}}_{\dot{2}-}$ and the combination $(\lambda^{12}_{-} ~-~ \lambda^{11}_{-})$ 
satisfy the same equations \eqref{fermeqs_ovpsi1_lplus_small-2} and are related to  solutions
\eqref{solution_orient_ov_psi_1-} and \eqref{solution_orient_lambda(1)} via
 \eqref{remainingferm}.

\subsection{Effective action in the orientational sector \label{superorient-eff_action}}

Now to see the effect of lifting of the orientational fermion zero modes 
let us derive  a fermionic part of the two-dimensional effective action on the string 
world sheet  with the $O(\mu)$ accuracy. In order to do so, we assume a slow $t$ and $z$ dependence
of the fermionic moduli $\xi^l_{L,R}$, 
substitute our {\it ansatz} \eqref{small_mu-supertransl-ansatz-lambda},  
\eqref{small_mu-supertransl-ansatz-psi}  into the four dimensional fermion action \eqref{fermact}
 and integrate over $x_1,\ x_2$. Kinetic terms for bulk fermions (containing derivatives
 $\p_0$ and $\p_3$) produce corresponding kinetic terms for two-dimensional fermions, 
and mass terms are generated because fermionic modes are now lifted. 
The result for the quadratic terms in the two dimensional fermionic action is
\begin{equation}
\mc{S}_{\rm 2d} ~~=~~ \int dt dz \,\left\{ \frac{4\pi}{g_2^2} (\bar{\xi}_L i\p_R \xi_L ~+~  \bar{\xi}_R i\p_L \xi_R) ~+~ m_{or }\gamma\, (\bar{\xi}_R \xi_L ~+~  \bar{\xi}_L \xi_R) + \cdots \right\},
\label{effective_action-superorient}
\end{equation}
where dots stand for higher order terms in fields and
\begin{equation}
\gamma ~~=~~ -4 \,  \int dx_1 dx_2\, \ov{\wt{\psi}}_{\dot{1}+} \, \ov{\psi}_{\dot{2}+} ~~=~~ 4\,  \int dx_1 dx_2\, |\ov{\psi}_{\dot{2}+}|^2 \, ,
\end{equation}
while
\[	
\p_R ~~=~~ \p_0 ~+~ i\, \p_3 ~, \qquad\qquad  \p_L ~~=~~ \p_0 ~-~ i\, \p_3~.
\]
We see that all two-dimensional fermionic fields $\xi^l_{L,R}$ become massive with mass
$m_{or}$ proportional to $\mu$. We expect that in the limit of large $\mu$ these fermions decouple
from bosonic CP$(N-1)$ model \eqref{cp}.

\subsection{Supertranslational zero modes}

As we already mentioned, supertranslational modes  
can be obtained via a supersymmetry transformation from bosonic string solution even in the 
$\mu$-deformed theory. String solution ceases to be a BPS one, and all of the four remaining supercharges 
$Q^{\alpha 1}$ of the \none theory act non-trivially on the string solution. Much in the same way as the 
bosonic translational modes, the supertranslational ones decouple from orientational CP$(N-1)$
model and are described by free fermions on the string world sheet. This can be anticipated
on general grounds.  To see this note that the orientational
fermion fields $\xi^l_{L,R}$ become heavy at large $\mu$ and without them  
we cannot construct interaction terms of $n^l$ and  supertranslational moduli $\zeta_{L,R}$
compatible with symmetries of the theory (if we do not consider higher derivative corrections). 

For the sake of completeness
we construct explicitly supertranslational zero modes in the large $\mu$ limit in Appendix B
acting by \none supersymmetry
transformations on the string solution of Sec.~\ref{sec:bosonic_solution}.

%
%

\section{Physics of the world sheet theory and confined monopoles
\label{physics}}
\setcounter{equation}{0}

As we have seen above the fermionic fields $\xi^l$ of the effective world sheet theory 
 become heavy in the large $\mu$ limit and decouple. Moreover, the translational sector
is free and does not interact with the orientational sector. Thus, our effective world sheet theory on the 
non-Abelian string is given by  bosonic 
CP($N$-1) model \eqref{cp} without fermions  in the large $\mu$ limit. If quark masses are 
small but not equal, the orientational moduli $n^l$ are lifted by shallow  potential \eqref{V1+1}.

As we already mentioned, our four dimensional bulk theory is in the Higgs phase where squarks
develop condensate \eqref{qVEV}. Therefore 't Hooft-Polyakov monopoles present in the  theory
in the \ntwo limit of small $\mu$ are confined by non-Abelian strings. In fact in U$(N)$ gauge theories
confined monopoles are junctions of two distinct  strings \cite{T,SYmon,HT2}. In the 
effective world sheet theory on the non-Abelian string they are seen as kinks interpolating between 
different vacua of CP$(N-1)$ model, see \cite{SYrev} for a review.

The question of the crucial physical importance is whether monopoles survive the limit
of large $\mu$ when the the bulk theory flows to \none QCD. Quasiclassically we do not expect this to happen.
From a quasiclassical point of view, the very existence of 't Hooft-Polyakov monopoles relies on the presence of
adjoint scalars which develop VEVs. At large $\mu$ adjoint fields become heavy and decouple in our 
bulk theory, so quasiclassically we do not expect monopoles to survive.

We will see now that in quantum theory the story becomes more interesting. Confined monopoles 
are represented by kinks of CP$(N-1)$ model on the non-Abelian string. Therefore to address the above
problem we have to study kinks in the world sheet theory. Certain results in this direction 
were already obtained. As we mentioned before, in the framework of massless $\mu$-deformed 
\ntwo QCD with FI $D$-term it was shown that the effective theory on the string world sheet is 
heterotic \ntwoo supersymmetric CP$(N-1)$ model \cite{Edalati,Tongd,SY02,BSYhet}. 
This model has $N$ degenerative vacua and kinks interpolating between them. This means that
kinks/confined monopoles do survive the large $\mu$ limit in the above mentioned theory.

In this paper we study a more ''realistic'' version of $\mu$-deformed theory without the FI $D$-term.
This theory flows to \none QCD in the large $\mu$-limit. As we have shown  the 
world sheet theory on the non-Abelian string reduces to non-supersymmetric CP$(N-1)$ model without
fermions in the large $\mu$ limit. 
If quark mass differences  are non-zero, a potential  \eqref{V1+1} is generated.
It does not have multiple local minima, therefore kinks (confined monopoles of the bulk theory) 
become unstable and disappear.

Consider the case when quark masses are equal. Then CP$(N-1)$ model is at strong coupling.
This model was solved by Witten \cite{W79} in the large $N$ approximation. It was shown that 
kinks in this model are in a confinement phase. In terms more suitable for application to monopole physics 
of the bulk theory this can be understood as follows, see also \cite{SYrev} for a more detail review.

\begin{figure}
\centering
\includegraphics[width=6cm]{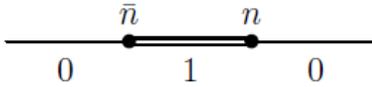}
\caption{Configuration of a string with kink and anti-kink on it. Zero and one represent 
the true vacuum and the first quasi-vacuum respectively.}
\label{diag3} 
\end{figure}

The vacuum structure of the $CP(N-1)$ model was studied in  \cite{W2}. It was shown that the genuine vacuum 
is unique. There are, however, of order $N$ quasi-vacua, which become stable in the 
limit $N\rightarrow\infty\,,$ since an energy split between the neighboring quasi-vacua 
is $O(1/N)$. Thus, one can imagine a kink interpolating between the true 
vacuum  and the first quasi-vacuum and the anti-kink returning to the true vacuum as 
in Fig.(\ref{diag3}). Linear confining potential between 
kink and anti-kink is associated with excited quasi-vacuum.

This two dimensional confinement of kinks was interpreted in terms of strings and monopoles of the bulk
theory in \cite{GSY05}. The fine structure of vacua in CP$(N-1)$ model on the non-Abelian string means that
$N$ elementary strings are split by quantum effects and  have slightly different tensions. The difference 
between the tensions of ''neighboring'' strings is proportional to $\Lambda_{CP}^2$, see \eqref{LambdaCP}.
Therefore monopoles, in addition to the four dimensional confinement (which ensures that
they are attached to the string), acquire two-dimensional confinement along the string. 
Monopole and antimonopole connected by a string with larger tension form a mesonic bound state.

Consider a monopole-antimonopole pair  interpolating between strings 0 and 1, see Fig.~\ref{diag3}. 
Energy of the excited part of the string (labeled as $1$) is proportional to the distance $R$ 
between the kink and anti-kink as 
\beq
V(R) \sim \Lambda_{CP}^2\,R.
\eeq
 When it exceeds the 
mass of  two monopoles which is of order of $\Lambda_{CP}$ then the second monopole-antimonopole pair 
emerges breaking the excited part of the string. This gives an estimate for the typical 
length of the excited part of the string, $R\sim N/\Lambda_{CP}$. Since this length grows in the large 
$N$ limit, kinks are metastable with an exponentially small decay rate $\exp{-N}$.

%
%

\section{Conclusions}
 \label{Conclusions}

In this paper we considered non-Abelian strings in \ntwo supersymmetric QCD deformed by a large mass 
term for adjoint matter. In the limit of large $\mu$ this theory flows to \none SQCD. We found a
solution for the non-Abelian string and derived a two dimensional effective theory on the string world sheet
which describes the dynamic of its orientational zero modes. This theory turns out to be 
bosonic CP$(N-1)$ model \eqref{cp} with shallow potential \eqref{V1+1} generated by small quark
mass differences. The fermionic superpartners $\xi^l$ of the bosonic orientational moduli $n^l$ present in 
the \ntwo limit become heavy at large $\mu$ and decouple.

We addressed the question of what happen to confined 't Hooft-Polyakov monopoles
at large $\mu$. We showed that, if the quark mass differences are larger than (exponentially small) $\Lambda_{CP}$,
the confined monopoles become unstable at large $\mu$. However, if 
the quarks have equal masses, the confined monopoles survive in the \none QCD limit. 
 This result is quite remarkable since \none QCD is in the non-Abelian regime and quasiclassically we 
do not expect monopoles in this theory. It also supports the picture of ''instead-of-confinement''
phase for \none QCD at strong coupling \cite{SYdualrev}.

\section*{Acknowledgments}

The authors are grateful to Mikhail Shifman   for very useful and
stimulating discussions.
The work of A.Y. was  supported in part by DOE grant DE-SC0011842, by William I. Fine Theoretical Physics Institute  of the  University
of Minnesota, and by Russian State Grant for
Scientific Schools RSGSS-657512010.2. The work of A.Y. was supported by Russian Scientific Foundation
under Grant No. 14-22-00281.


\section*{Appendix A:}
\addcontentsline{toc}{section}{Appendix A}
\renewcommand{\theequation}{A.\arabic{equation}}
\setcounter{equation}{0}

In this Appendix we solve Dirac equations \eqref{fermeqs_ovpsi1_lplus_small-1}.
After a substitution
\begin{align}
\notag
\ov{\psi}_{\dot{1}-}(r) &~~=~~ \frac{1}{r\phi_2(r)}\Psi(r),\\
\notag
\lambda_{(1)}(r) &~~=~~ i g_2^2 \Lambda(r)
\end{align}
equations \eqref{fermeqs_ovpsi1_lplus_small-1} reduce to
\begin{equation}
\begin{aligned}
\frac{1}{r g_2^2 \phi_1\phi_2} \p_r \Psi &~~=~~ \Lambda \,,\\[2mm]
r\p_r\Lambda ~+~ f_N\Lambda
~-~ \frac{\phi_1}{\phi_2}\Psi &~~=~~  ~-~  \frac{\mu_2 f_N}{2} \frac{\phi_1}{\phi_2} \, ,
\end{aligned}
\label{newvariables}
\end{equation}
which in turn gives an equation of second order for $\Psi$:
\begin{equation}
\p_r^2 \Psi ~-~ \frac{1}{r}\left( 1 ~+~ \frac{2}{N}(f-f_N) \right)\, \p_r \Psi ~-~ g_2^2\phi_1^2\Psi ~~=~~ -\,\frac{\mu_2 f_N}{2} g_2^2\phi_1^2 \,.
\label{fermeqs_Psi_small}
\end{equation}
First let us solve the homogeneous version of \eqref{fermeqs_ovpsi1_lplus_small-1}, i. e. 
put $\mu_2 ~=~ 0$. The solutions are
\begin{align*}
\ov{\psi}_{\dot{1}-} ~~=~~ c\frac{f_N}{r\phi_2}, \\[2mm]
\lambda_{(1)} ~~=~~ c\frac{i g_2^2}{2} \left( \frac{\phi_1}{\phi_2} ~-~ \frac{\phi_2}{\phi_1} \right).
\end{align*}
with some constant $c$. %
They correspond to $\Psi ~=~ f_N$; indeed, this is a solution for homogeneous version of \eqref{fermeqs_Psi_small}. With the help of it we can reduce the order of this equation. Let us take
\[
\Psi(r) ~~=~~ \mu_2\, f_N(r)\left( \int\limits_0^r {\mathrm d}x\, \chi(x) + c_1 \right),
\]
with some constant $c_1$, then from \eqref{fermeqs_Psi_small} it follows that
\begin{equation}
\p_r \chi ~+~ \frac{1}{r}\left( \frac{1}{f_N}r^2g_2^2(\phi_1^2 ~-~ \phi_2^2) ~-~ 1 ~-~ \frac{2}{N}(f-f_N) \right)\, \chi ~~=~ -\,\frac{1}{2} g_2^2  \phi_1^2 \,.
\label{fermeqs_theta_small}
\end{equation}
This is just an equation of the first order; its solution can be found very easily as
\begin{equation}
\chi ~~=~~ -\, \frac{g_2^2 r \phi_2^2}{2 f_N^2} \left(  \int\limits_0^r \frac{{\mathrm d}y}{y}\, \frac{\phi_1^2}{\phi_2^2} f_N^2 ~+~ c_2 \right).
\end{equation}
Putting all this together, we obtain:
\begin{equation}
\ov{\psi}_{\dot{1}-}(r) ~~=~~ -\,\mu_2\, g_2^2\, \frac{f_N(r)}{r\phi_2(r)} \left( \int\limits_0^r {\mathrm d}x\,
\frac{ x \phi_2^2(x)}{2 f_N^2(x)} \left(  \int\limits_0^x \frac{{\mathrm d}y}{y}\, \frac{\phi_1^2(y)}{\phi_2^2(y)}f_N^2(y) ~+~ c_2 \right) + c_1 \right)   .
\label{fermsol_ovpsi1_lplus_small-const}
\end{equation}
with some new constant $c_1$.

For this solution to behave well at the origin we have to put $c_1 ~=~ 0$. Considering the infinity, we should also require that
\[
c_2 ~~=~~ -\, \int\limits_0^\infty \frac{{\mathrm d}y}{y}\, \frac{\phi_1^2(y)}{\phi_2^2(y)}f_N^2(y).
\]
This gives
\begin{equation}
\ov{\psi}_{\dot{1}-}(r) ~~=~~ \frac{\mu_2\, g_2^2}{2}\, \frac{f_W(r)}{r\phi_2(r)}  \int\limits_0^r {\mathrm d}x\,
\frac{ x \phi_2^2(x)}{f_W^2(x)}   \int\limits_x^\infty \frac{{\mathrm d}y}{y}\,
 \frac{\phi_1^2(y)}{\phi_2^2(y)}f_W^2(y)   .
\label{fermsol_ovpsi1_lplus_small}
\end{equation}
for for $\ov{\psi}_{\dot{1}-}$ and
\begin{equation}
\begin{multlined}
\lambda_{(1)}(r) ~~\equiv~~ \lambda^{22}_- ~+~ \lambda^{21}_- ~~=~~ \\
\phantom{\lambda^{22}_- ~+~ \lambda^{21}_-} =~~ \frac{i\, \mu_2\, g_2^2}{2}\, \Bigg( \frac{g_2^2}{2}
\left( \frac{\phi_1}{\phi_2} ~-~ \frac{\phi_2}{\phi_1} \right)
\int\limits_0^r {\mathrm d}x\, \frac{ x \phi_2^2(x)}{f_W^2(x)}   \int\limits_x^\infty \frac{{\mathrm d}y}{y}\, \frac{\phi_1^2(y)}{\phi_2^2(y)}f_W^2(y) \\
~+~ \frac{\phi_2}{\phi_1 f_W} \int\limits_r^\infty \frac{{\mathrm d}y}{y}\, \frac{\phi_1^2(y)}{\phi_2^2(y)}f_W^2(y)
\Bigg) .
\end{multlined}
\label{fermsol_lamplus_lplus_small}
\end{equation}
 for  $\lambda_{(1)}$.
By direct substitution we  verified that these modes indeed satisfy the Dirac equations.

Now let us consider Dirac equations \eqref{fermeqs_ovpsi1_lplus_small-2} with the non-zero eigenvalue 
$m_{or}$. Applying the method developed above we find the solutions
\begin{equation}
\ov{\psi}_{\dot{1}-}(r) ~~=~~ -\left(m_{or } - \mu_2\frac{g_2^2}{2} \right)\,\frac{f_N(r)}{r\phi_2(r)}\int\limits_0^r {\mathrm d}x \frac{x\phi_2^2(x)}{f_N^2(x)}  \int\limits_x^\infty {\mathrm d}y \frac{f_N^2(y)\phi_1^2(y)}{y\phi_2^2(y)}
\label{solution_orient_ov_psi_1-}
\end{equation}
for $ \ov{\psi}_{\dot{1}-}$ and  
\begin{equation}
\begin{multlined}
\lambda_{(1)}(r) ~~=~~ 
 -i\frac{m_{or }}{2}\left( \frac{\phi_1}{\phi_2} ~-~ \frac{\phi_2}{\phi_1}  \right)
~-~
ig_2^2\left(m_{or } - \mu_2\frac{g_2^2}{2} \right)\,\frac{1}{2}\left( \frac{\phi_1}{\phi_2} ~-~ 
\frac{\phi_2}{\phi_1}  \right)\, \times \\
\int\limits_0^r {\mathrm d}x \frac{x\phi_2^2(x)}{f_N^2(x)}  
\int\limits_x^\infty {\mathrm d}y \frac{f_N^2(y)\phi_1^2(y)}{y\phi_2^2(y)}
~-~
i\left(m_{or } - \mu_2\frac{g_2^2}{2} \right)\,\frac{\phi_2}{f_N\phi_1}
\int\limits_r^\infty {\mathrm d}y \frac{f_N^2(y)\phi_1^2(y)}{y\phi_2^2(y)}
\end{multlined}
\label{solution_orient_lambda(1)}
\end{equation}
 for $\lambda_{(1)}$.

One can see, that the first and the last terms in the last expression behave at the origin as $1/r$. 
We can choose eigenvalue $m_{or }$ to insure that $1/r$ terms cancel out. This gives the expression
\eqref{solution_orient_alpha} for the mass $m_{or }$.

\section*{Appendix B:}
\addcontentsline{toc}{section}{Appendix B}
\renewcommand{\theequation}{B.\arabic{equation}}
\setcounter{equation}{0}

In this Appendix we generate supertranslational fermion zero modes in the large $\mu$ limit 
acting by \none supersymmetry
transformations on the string solution of Sec.~\ref{sec:bosonic_solution}. The \none supersymmetry
transformations have the form
\begin{eqnarray}
\delta\bar{\tilde\psi}_{\dot{\alpha}}^{kA}
&=&
i\sqrt2\
\bar\nabla\hspace{-0.65em}/_{\dot{\alpha}\alpha}q^{kA}\epsilon^{\alpha } ~+~ \sqrt{2}\, \bar{\epsilon}^{\alpha }\bar{\tilde F}^{kA}  \  \ ,
\nonumber\\[4mm]
\delta\bar\psi_{\dot{\alpha}Ak}
&=&
i\sqrt2\
\bar\nabla\hspace{-0.65em}/_{\dot{\alpha}\alpha}\bar{q}_{\,Ak}\epsilon^{\alpha } ~+~ \sqrt{2}\, \bar{\epsilon}^{\alpha }\bar{F}_{kA}  \ ,
\label{transf_n=1}
\end{eqnarray}
where the $F$-terms
are given by derivatives of the superpotential \eqref{superpotential-adjoints_integrated}, 
\begin{equation}
\begin{aligned}
\bar{F}_{Ak} ~~=~~ -\frac{\p \mc{W}}{\p q^{kA}} &~=~ \frac{i}{\mu_2} \left( \tilde{q}_{Ck} (\tilde{q}_A q^C) - \frac{\alpha}{N} (\tilde{q}_C q^C)\tilde{q}_{Ak} \right) + m \tilde{q}_{Ak} \\
&~=~ \frac{i}{\mu_2} \left( \vp_{Ck} (\vp_A \vp^C) - \frac{\alpha}{N} (\vp_C \vp^C)\vp_{Ak} \right) + m \vp_{Ak} \,,
\end{aligned}
\end{equation}
\begin{equation}
\begin{aligned}
\bar{\tilde F}^{kA} ~~=~~ -\frac{\p \mc{W}}{\p \tilde{q}_{Ak}} &~=~ \frac{i}{\mu_2} \left( q^{kC} (\tilde{q}_C q^A) - \frac{\alpha}{N} (\tilde{q}_C q^C)q^{kA} \right) + m q^{kA} \\
&~=~ \frac{i}{\mu_2} \left( \vp^{kC} (\vp_C \vp^A) - \frac{\alpha}{N} (\vp_C \vp^C)\vp^{kA} \right) + m \vp^{kA}  \,,
\end{aligned}
\end{equation} 
where we also used \eqref{q-ansatz}.

Consider first the region of intermediate $r$,
 in the 
range $1/m_G \lesssim r \lesssim 1/m_L$. 
As we will see, the fermion zero modes  behave as $1/r$. This will give us leading logarithmic 
contributions to the kinetic terms for fermions of two dimensional
effective theory on the string world sheet. 

To calculate the fermionic modes one should substitute bosonic solutions
 \eqref{string:boson:large_mu:dm=0:intermediate} into the 
transformations \eqref{transf_n=1}.  In \eqref{transf_n=1}, the first terms in the first and second 
  lines give $1/r$ contributions, whereas the $F$-terms adds  constant and logarithmic terms, which 
does not produce leading logarithmic terms in the effective action. 
 We neglect these last terms, and get non-zero fermionic 
profiles 
%
%
%
\begin{equation}
\begin{aligned}
\ov{\psi}{}_{\dot{1}} & ~~\approx~~   (n\nbar)\ \frac{\displaystyle x^1 - i\, x^2}{\displaystyle r}  \frac{1}{r} \ \frac{\sqrt{\xi}}{\ln \frac{g_2\sqrt{\xi}}{m_L}} \, \zeta_R
~, \\[2mm]
\ov{\psi}{}_{\dot{2}} & ~~\approx~~  (n\nbar)\ \frac{\displaystyle x^1 + i\, x^2}{\displaystyle r}  \frac{1}{r} \ \frac{\sqrt{\xi}}{\ln \frac{g_2\sqrt{\xi}}{m_L}} \, \zeta_L
~, \\[2mm]
\ov{\wt{\psi}}{}_{\dot{1}} & ~~\approx~~ (n\nbar)\ \frac{\displaystyle x^1 - i\, x^2}{\displaystyle r}  \frac{1}{r} \ \frac{\sqrt{\xi}}{\ln \frac{g_2\sqrt{\xi}}{m_L}}  \, \zeta_R
~, \\[2mm]
\ov{\wt{\psi}}{}_{\dot{2}} & ~~\approx~~ (n\nbar)\ \frac{\displaystyle x^1 + i\, x^2}{\displaystyle r}  \frac{1}{r} \ \frac{\sqrt{\xi}}{\ln \frac{g_2\sqrt{\xi}}{m_L}}  \, \zeta_L .
\end{aligned}
\label{supertranslational_N=1_psi_solutions}
\end{equation}
One can see that these modes are indeed proportional to $1/r$.
Here $\zeta_{L,R}$ are the Grassmann parameters generated by supersymmetry transformations,
\mbox{$\zeta_L = \frac{1}{\sqrt{2}}\epsilon^{1}, \  \zeta_R = -\frac{1}{\sqrt{2}} \epsilon^{2}$}).
These parameters become fermionic fields in the two dimensional effective 
theory on the string world sheet.

The region of small $r$, $r\ll 1/m_G$ does not contribute because 
quark fields vanish in this limit.

To find  the effective world sheet action, one should substitute  solutions \eqref{supertranslational_N=1_psi_solutions} into four-dimensional fermionic action \eqref{fermact}. For the kinetic term, we obtain:
\begin{equation}
\mc{L}_{\rm 2d} ~~=~~ 2\pi\xi \, I_\xi (\bar{\zeta}_L i\p_R \zeta_L ~+~  \bar{\zeta}_R i\p_L \zeta_R) \,,
\label{action:supertrans_2d}
\end{equation}
where the normalization constant is
\begin{equation}
I_\xi ~~=~~ \frac{2N}{\ln{\frac{m_W}{m_L}}} \,.
\end{equation}
As we already mentioned,  deriving this effective action we integrated over the transversal coordinates in the 
range $1/m_G \lesssim r \lesssim 1/m_L$. The integral over $r$ is logarithmically enhanced.
 The contributions of other regions do not have  logarithmic enhancement and can be 
neglected.

We see that \eqref{action:supertrans_2d} is the action for free fermions which decouples from
the orientational sector given by CP$(N-1)$ model \eqref{cp}. 
%
%



\end{document}